\documentclass[a4paper,11pt]{article}
\pdfoutput=1 

\usepackage{jcappub} 

\usepackage[T1]{fontenc} 
\usepackage[utf8]{inputenc}
\usepackage{siunitx}
\usepackage{bm}
\usepackage{dsfont}

\title{Fast Point Spread Function Modeling with Deep Learning}

\author[a]{J\"org Herbel,}
\author[a]{Tomasz Kacprzak,}
\author[a]{Adam Amara,}
\author[a]{Alexandre Refregier,}
\author[b]{Aurelien Lucchi}

\affiliation[a]{Institute for Particle Physics and Astrophysics, Department of Physics, ETH Zurich, Wolfgang-Pauli-Strasse 27, CH-8093 Zurich, Switzerland}
\affiliation[b]{Data Analytics Lab, Department of Computer Science, ETH Zurich, Universit\"atstrasse 6, CH-8092 Zurich, Switzerland}

\emailAdd{joerg.herbel@phys.ethz.ch}
\emailAdd{tomasz.kacprzak@phys.ethz.ch}
\emailAdd{adam.amara@phys.ethz.ch}
\emailAdd{alexandre.refregier@phys.ethz.ch}
\emailAdd{aurelien.lucchi@inf.ethz.ch}

\abstract{Modeling the Point Spread Function (PSF) of wide-field surveys is vital for many astrophysical applications and cosmological probes including weak gravitational lensing. The PSF smears the image of any recorded object and therefore needs to be taken into account when inferring properties of galaxies from astronomical images. In the case of cosmic shear, the PSF is one of the dominant sources of systematic errors and must be treated carefully to avoid biases in cosmological parameters. Recently, forward modeling approaches to calibrate shear measurements within the Monte-Carlo Control Loops (\textit{MCCL}) framework have been developed. These methods typically require simulating a large amount of wide-field images, thus, the simulations need to be very fast yet have realistic properties in key features such as the PSF pattern. Hence, such forward modeling approaches require a very flexible PSF model, which is quick to evaluate and whose parameters can be estimated reliably from survey data. We present a PSF model that meets these requirements based on a fast deep-learning method to estimate its free parameters. We demonstrate our approach on publicly available SDSS data. We extract the most important features of the SDSS sample via principal component analysis. Next, we construct our model based on perturbations of a fixed base profile, ensuring that it captures these features. We then train a Convolutional Neural Network to estimate the free parameters of the model from noisy images of the PSF. This allows us to render a model image of each star, which we compare to the SDSS stars to evaluate the performance of our method. We find that our approach is able to accurately reproduce the SDSS PSF at the pixel level, which, due to the speed of both the model evaluation and the parameter estimation, offers good prospects for incorporating our method into the \textit{MCCL} framework.}

\begin{document}
\maketitle
\flushbottom

\section{Introduction}

The Point Spread Function (PSF) of an observation describes the mapping of a point source in the sky onto the two-dimensional image plane (e.g. \cite{Bradt2003}). Diffraction effects in the optical components of the telescope as well as turbulences in the Earth's atmosphere (for ground-based facilities) cause a point source to be mapped into an extended object with a finite size. Other effects such as imperfections in optical elements along the light path and charge diffusion in digital detectors also contribute to the PSF, which affects all observed objects by smearing their images. Mathematically, this can be described as a convolution of the intrinsic light distribution with a kernel characterizing the PSF. Point sources, such as stars, are effectively noisy samples of the PSF at the position of the star. Extended objects like galaxies appear larger due to the smearing caused by the PSF.

The PSF impacts any measurement of galaxy properties from astronomical images. Therefore, estimating and modeling the PSF is important in many branches of astrophysics and cosmology. For example, the PSF needs to be taken into account when studying galaxy morphology by fitting observed objects with models of the light distribution (e.g. \cite{Haeussler2007, Gabor2009}). Another example is the analysis of strong lensing systems \cite{Birrer2016, Wong2017}. When modeling the measured light distribution, the PSF has to be included in the model to avoid biased results. Also, cosmic shear measurements (reviewed in \cite{Refregier2003b, Hoekstra2008, Kilbinger2015}) require a precise treatment of the PSF, which induces distortions that can contaminate the measured lensing signal (e.g. \cite{Heymans2012}).

The impact of incorrectly modeling the PSF on weak lensing measurements has been studied extensively (for example \cite{Paulin-Henriksson2008, Paulin-Henriksson2009, Amara2010, Massey2013}). This is because the PSF is one of the dominant sources of systematic errors for cosmic shear and thus needs to be taken into account carefully to avoid biases in the cosmological parameter inference. Various methods have been developed to model the PSF for weak lensing analyses. For example, the cosmic shear pipeline of the Kilo-Degree Survey (KiDS) \cite{Hildebrandt2017} uses shapelets \cite{Refregier2003a} as basis functions to model the PSF. Another option, pursued by the weak lensing pipeline of the Dark Energy Survey (DES) \cite{Zuntz2017}, is a pixel-based approach \cite{Bertin2011}. In this case, a basis for PSF modeling is obtained directly from the data by performing a principal component analysis (reviewed in \cite{Jollife2016}). Further methods to estimate and model the PSF are given in \cite{Kitching2013}.

In recent years, forward modeling has become an increasingly popular method to calibrate shear measurements. A consistent framework for this is the Monte-Carlo Control Loops (\textit{MCCL}) scheme \cite{Refregier2014, Bruderer2016, Herbel2017, Bruderer2017}, originally developed for wide-field imaging data. This method typically requires simulating a large amount of synthetic images with realistic properties. One of the most important features that the simulations need to capture is the PSF pattern in the data, since an incorrectly modeled PSF will lead to model biases in the measured lensing signal. To do this, the PSF model needs to be flexible so as to capture possible complexities in the data while being fast enough to run on large datasets.

In this paper, we construct a PSF model that is informed by data. We do this using a principal component analysis of the stars in our sample, which informs us about the main features our model has to reproduce. The model consists of a base profile that is distorted through a pertubative expansion. This model has several parameters that we fit using a machine-learning method known as Convolutional Neural Network (CNN). Deep-learning methods such as CNNs have been shown to perform especially well on imaging data (e.g. \cite{LeCun2015}). An additional reason for choosing a CNN is that these networks are very fast once trained, which is crucial in the context of forward modeling and \textit{MCCL}. Deep Learning has gained increasing attention in astrophysics and cosmology recently and it has been applied successfully to various problems in the field (see e.g. \cite{Ravanbakhsh2016, Charnock2017, Hezaveh2017, Schmelzle2017, Lanusse2018}). Specifically, we use a deep network to solve a regression problem, namely to map the noisy image of a star to the parameter space of our model, allowing us to simulate the PSF at that position. We test and demonstrate our method using publicly available data from the Sloan Digital Sky Survey (SDSS).

The paper is structured as follows: In section \ref{sec:sdss-star-sample}, we introduce the data used to demonstrate our method. Section \ref{sec:psf-model} describes our model and the data-based approach we pursue to construct it. In section \ref{sec:cnn}, we present our deep-learning approach to estimate the parameters of our model from images of stars. Our results are shown in section \ref{sec:results} and we conclude with section \ref{sec:conclusion}.

\section{\label{sec:sdss-star-sample}SDSS star sample}

We use publicly available SDSS $r$-band imaging and catalog data to obtain a set of stars on which we demonstrate the applicability of our method. We query the SDSS database\footnote{\url{http://skyserver.sdss.org/CasJobs/}} for all  $r$-band images from the SDSS data release 14 \cite{Abolfathi2017} located between \ang{100} and \ang{280} right ascension and \ang{0} and \ang{60} declination. From this set of images, we select the ones with at least one object that fulfills the following criteria: (i) the object was classified as a star in the $r$-band by the SDSS pipeline, (ii) the object has clean photometry, (iii) the object has an $r$-band signal-to-noise ratio of at least 50. The corresponding query submitted to the SDSS database is detailed in appendix \ref{app:casjobs-sql-query}. This leads to $248\,866$ images and corresponding catalogs.

To obtain a high-purity sample of stars from this data, we additionally match the SDSS catalogs to the stars found in the \textit{Gaia} data release 1\footnote{\url{https://gea.esac.esa.int/archive/}} \cite{GaiaCollaboration2016a, GaiaCollaboration2016b}. We cross-match sources in \textit{Gaia} and SDSS based on positions. To be considered a match, a source in SDSS must be closer than $2''$ to a source in the \textit{Gaia} catalog. Additionally, we ensure that there are no other sources in SDSS within a radius of $8.5''$. This criterion guarantees that we only use stars without close neighbors, which would contaminate the images of the objects in our sample.

After defining our sample of stars, we produce $15 \times 15$ pixel cutouts centered on the pixel with the maximum flux. Additionally, at this step, we remove objects that are too close to the border of the SDSS image so that we cannot make a complete stamp, i.e. the peak flux is within 7 pixels of the border. This results in $1\,215\,818$ cutouts. We normalize each cutout to have a maximum pixel intensity of 1, so that all stamps cover similar numerical ranges. This step removes the flux information from the cutouts, which will allow our CNN to process a wide range of stars with fluxes that vary by several orders of magnitude. A random selection of six normalized cutouts is shown in figure \ref{fig:sdss-stars--vs--cnn-stars}.

\section{\label{sec:psf-model}PSF model}

\subsection{Model construction}

To construct a model of the PSF images in our sample, we need to know about the features this model has to capture. We therefore need to find the main modes of variation in our data, which will allow us to incorporate these characteristics into our model. One approach to this problem is a principal component analysis (PCA, see \cite{Jollife2016} for a review), which yields a set of basis vectors our data decomposes into. These basis vectors, called principal components, are ordered according to the amount of variation they introduce in our sample. Therefore, a reasonable reconstruction of our data can be achieved using the first few principal components only, since they account for the most prominent features. For this reason, a PCA is well-suited for building a model that captures the most important characteristics of our data.

\begin{figure}[tbp]
\centering
\includegraphics[width=\textwidth]{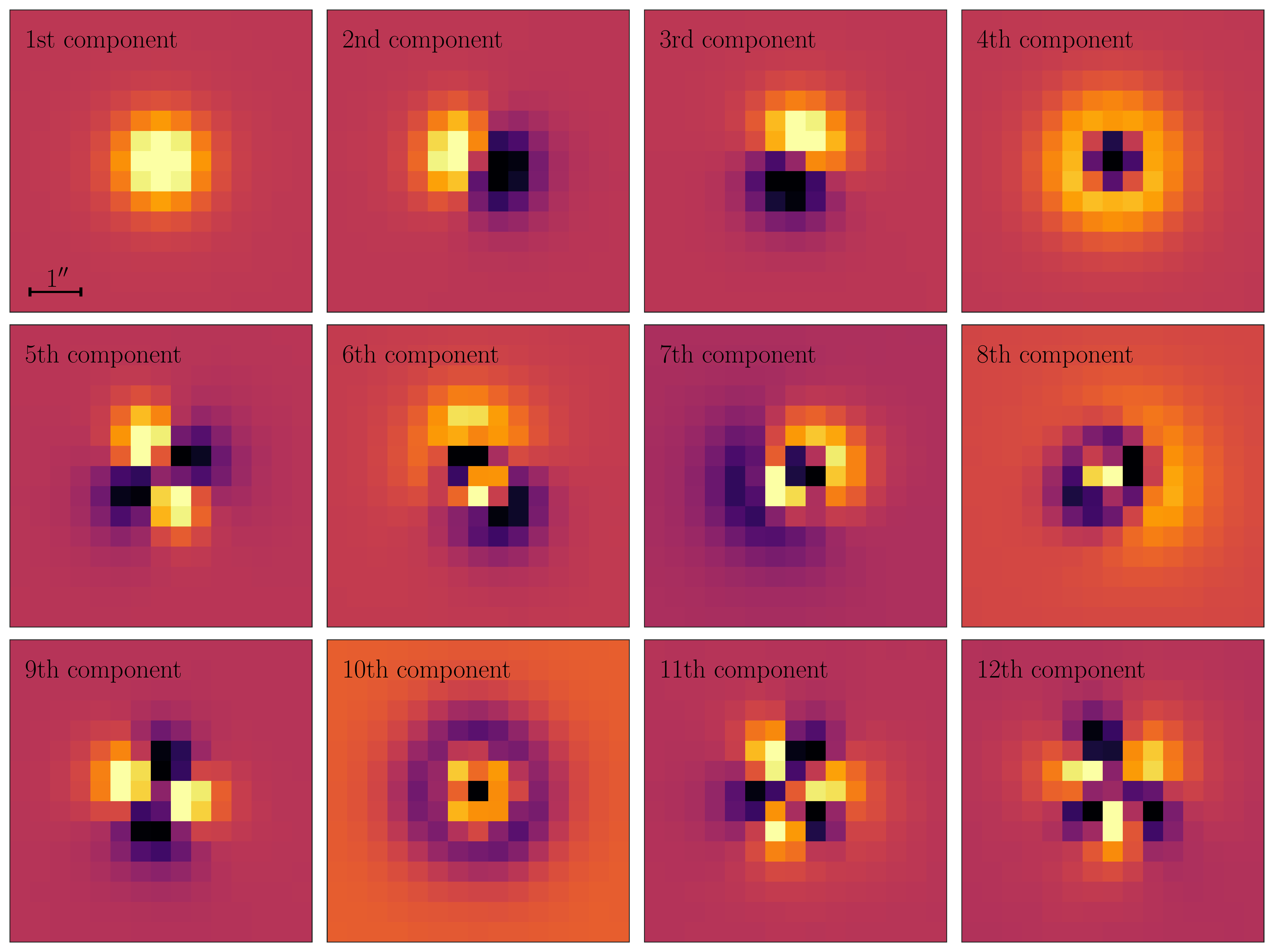}
\caption{\label{fig:svd-sdss}The first 12 (uncentered) principal components of our SDSS sample, which have been obtained by performing a SVD. The first component is the pixel-wise mean SDSS star. The second and third component indicate shifts of the positions of the centers of the stars. The fourth component arises due to variations in the size of the PSF and the quadrupoles seen in the components 5 and 9 indicate variations in ellipticity. The components 6, 7, 11 and 12 indicate the presence of flexion in the light distributions of the SDSS stars, while the 10th component represents kurtosis. A more detailed description of the distortion effects seen in the principal components is given in section \ref{subsec:model-description}. The black reference bar denotes an angular distance of $1''$ (average value computed from all used SDSS images). Note that the color scale is the same for all panels.}
\end{figure}

To find the principal components of our sample, we perform a singular-value decomposition (SVD, \cite{Press2007}). We reshape our data into a two-dimensional array $\bm{S}$, whereby each row contains a flattened version of one cutout. Thus, $\bm{S}$ has dimensions $n_\text{star} \times n_\text{pix}$, where $n_\text{star} = 1\,215\,818$ is the number of stars in our sample and $n_\text{pix} = 15^2$ is the number of pixels in one cutout. We then apply the SVD to $\bm{S}$ and reshape the resulting principal components to the original shape of $15 \times 15$ pixels. Note that we do not center our data before performing the SVD, thus we actually compute uncentered principal components. This is intended, since we aim at extracting a basis set that describes the data in terms of perturbations of the mean SDSS PSF image. The reason for this will become clear further below, where we describe our approach to constructing a model of the SDSS PSF based on the results of the SVD. Figure \ref{fig:svd-sdss} shows the first 12 uncentered principal components of our sample. The corresponding singular values drop by two orders of magnitude over this range.

It is clear from figure \ref{fig:svd-sdss} that our sample decomposition resembles noisy, pixelated versions of the polar shapelets introduced in \cite{Refregier2003a, Massey2005}. These basis functions can be used to analyze and decompose localized images and allow for an intuitive interpretation of the decomposition. We make use of the fact that our sample decomposes into shapelets by constructing our PSF model in the following way. First, we choose a base profile that we keep constant for all stars. We then introduce a set of distortion operations and a set of corresponding distortion parameters that allow for deviations from the base profile. These parameters vary from star to star to capture variations of the PSF. Due to our sample decomposing into shapelets, we know which kind of distortion operations we have to apply to our base profile in order to reproduce the characteristics seen in figure \ref{fig:svd-sdss}. In the following section, we elaborate on the details of our model. Note that our model is different from shapelets, since we use a non-gaussian base profile.

\subsection{\label{subsec:model-description}Model description}

Our PSF model is fully described by two components: a base profile that is kept constant for all stars and a set of distortion parameters that captures deviations from the base profile. These parameters vary from star to star to allow for variations in the simulated PSF images. The base profile accounts for the first component seen in figure \ref{fig:svd-sdss}, while the perturbations account for the other components. Simulating a star with this model is done in two steps. First, we draw photons from the base profile. In the second step, we distort the photon positions, inducing the characteristics seen in figure \ref{fig:svd-sdss}.

We adopt the mixture of two Moffat profiles \cite{Moffat1969} as our radially symmetric base profile $I$:
\begin{align}
I(r) &= I_1(r) + I_2(r), \\ 
I_i(r) &= I_{0, i} \left[ 1 + \left( \frac{r}{\alpha_i} \right)^2 \right]^{-\beta_i},
\end{align}
where $r$ is the distance from the center of the profile. The parameters $\beta_i$ set the concentrations of the profiles, while the $\alpha_i$ set the spatial spreads. The parameters $I_{0, i}$ determine the numbers of photons sampled from $I_i$. While the total number of photons for a given star, i.e. the flux, is allowed to vary, we keep the ratio $\gamma = I_{0, 1} / I_{0, 2}$ constant for all stars. Furthermore, we also set $\beta_i$ and $\alpha_i$ to the same values for all stars, which means that initially, all simulated objects are circularly symmetric and have the same size, namely a half-light radius $r_{50}$ of one pixel. This can be used to write $\alpha$ as 
\begin{equation}
\alpha = \frac{r_{50}}{\sqrt{2^{1 / (\beta - 1)} - 1}} = \frac{1}{\sqrt{2^{1 / (\beta - 1)} - 1}}.
\end{equation}
To allow for deviations from this base profile, we introduce a set of distortion operations that are applied to the photons after they were sampled from the base profile. In the following, we detail these operations and explain the perturbations they introduce into our base profile. This connects our model to the principal components displayed in figure \ref{fig:svd-sdss}.

\subsubsection{Size}

Variations in the size of the PSF are represented by the fourth principal component, which shows the type of residual that occurs in the case where the size is modeled incorrectly. We parameterize the size in terms of the full width at half maximum (FWHM), whereby we choose to use the same FWHM for $I_1$ and $I_2$. To simulate a star with a given FWHM $F$, we scale the positions of all photons by the corresponding half-light radius, which can be obtained via
\begin{equation}
r_{50, i} = \frac{\sqrt{2^{1 / (\beta_i - 1)} - 1}}{2 \sqrt{2^{1 / \beta_i} - 1}} \, F.
\end{equation}
Let $\vec{\theta} = (\theta_1, \theta_2)$ be the position of a photon sampled from the base profile. The distorted position $\vec{\theta}'$ resulting from a change in size is given by
\begin{equation}
\label{eq:size-transformation}
\theta'_i = r_{50} \, \theta_i.
\end{equation}

\subsubsection{Ellipticity, flexion and kurtosis}

The principal components 5 to 12 indicate the presence of ellipticity, flexion and kurtosis in the light distributions of the stars in our sample. We implement these distortions by applying a second transformation following after the first one given by eq. \eqref{eq:size-transformation}:
\begin{equation}
\theta''_i = A_{ij} \theta'_j + D_{ijk} \theta'_j \theta'_k + E_{ijkl} \theta'_j \theta'_k \theta'_l.
\end{equation}
This transformation is an expansion describing the distorted coordinates $\vec{\theta}''$ in terms of the coordinates $\vec{\theta}'$, which only include changes in the size. The transformation tensors $\bm{A}, \bm{D}, \bm{E}$ can be expressed in terms of derivatives of $\vec{\theta}''$ with respect to $\vec{\theta}'$:
\begin{equation}
\label{eq:transformation-derivatives}
A_{ij} = \frac{\partial \theta''_i}{\partial \theta'_j},  \qquad D_{ijk} = \frac{1}{2} \frac{\partial^2 \theta''_i}{\partial \theta'_j \partial \theta'_k}, \qquad E_{ijkl} = \frac{1}{6} \frac{\partial^3 \theta''_i}{\partial \theta'_j \partial \theta'_k \partial \theta'_l}.
\end{equation}
The same formalism is used to describe the effects of weak gravitational lensing, see \cite{Bacon2006}. In the following, we describe $\bm{A}$, $\bm{D}$ and $ \bm{E}$ in detail:
\begin{itemize}
\item The first-order term in the expansion of $\theta'_i$, $\bm{A}$, accounts for the ellipticity of the PSF, which is represented by the quadrupoles seen in the 5th and the 9th principal component. $\bm{A}$ can be parameterized by two parameters $e_1$ and $e_2$ (see e.g. \cite{Rhodes2000, Berge2013}) via 
\begin{align}
A &= \begin{cases} \frac{1}{\sqrt{2}} \begin{pmatrix} \text{sgn}(e_2) \, \sqrt{(1 + |e|)(1 + e_1 / |e|)} & - \sqrt{(1 - |e|)(1 - e_1 / |e|)} \\ 
                                                                                 \sqrt{(1 + |e|) (1 - e_1 / |e|)} & \text{sgn}(e_2) \, \sqrt{(1 - |e|) (1 + e_1 / |e|)} \end{pmatrix}, & |e| > 0, \\
                                                    \mathds{1}, & |e| = 0,
         \end{cases}
\end{align}
where $|e| = \sqrt{e_1^2 + e_2^2}$.
\item The second-order term $\bm{D}$ induces skewness (6th and 7th principal component) and triangularity (11th and 12th principal component) into the simulated light distribution. These two types of distortions are collectively referred to as flexion \cite{Bacon2006}. They can be parameterized by four variables $f_1$, $f_2$, $g_1$, $g_2$ according to:
 \begin{align}
 D_{ijk} &= \mathcal{F}_{ijk} + \mathcal{G}_{ijk}, \\ 
\mathcal{F}_{ij1} &= -\frac{1}{2} \, \begin{pmatrix} 3 f_1 & f_2 \\ f_2 & f_1 \end{pmatrix}, \quad  
\mathcal{F}_{ij2} = -\frac{1}{2} \, \begin{pmatrix} f_2 & f_1 \\ f_1 &  3 f_2 \end{pmatrix}, \\
\mathcal{G}_{ij1} &= -\frac{1}{2} \, \begin{pmatrix} g_1 & g_2 \\ g_2 &  -g_1 \end{pmatrix}, \quad
\mathcal{G}_{ij2} = -\frac{1}{2} \, \begin{pmatrix} g_2 & -g_1 \\ -g_1 &  -g_2 \end{pmatrix},                                                        
\end{align}
where $f_1$ and $f_2$ account for the skewness of the PSF, while $g_1$ and $g_2$ induce triangularity into our simulated light distributions.
\item We include one third-order term in our model, namely the one that accounts for kurtosis (symmetrically in all directions). This is motivated by the 10th principal component in figure \ref{fig:svd-sdss}, which displays the type of residual that arises from moving photons from the bulk of the distribution to the tail or vice-versa. The corresponding transformation that applies a kurtosis $k$ is given by
\begin{equation}
\theta''_i = k \, \theta'_i \, \left| \vec{\theta}' \right|^2,
\end{equation}
such that in our case, $E_{ijmn} = \delta_{ij} \, \delta_{mn} \, k$ for a given kurtosis $k$. 
\end{itemize}

\subsubsection{Centroid position}

Finally, we account for shifts of the position of the center of the profile, given by the second and third component in figure \ref{fig:svd-sdss}. This is achieved by offsetting all photon positions by a constant amount:
\begin{equation}
\theta'''_i = \theta''_i + \Delta \theta_i,
\end{equation}
whereby $\Delta \theta_i$ denotes a constant offset.

\subsubsection{Summary}
To summarize, we construct our PSF model in a data-driven way, which is motivated by the principal components our star sample decomposes into. To simulate a star, we first sample photons from a fixed base profile given by a weighted sum of two Moffat profiles. To perform the random draws, we adopt the sampling procedure detailed in \cite{Berge2013}. This results in the photon position $\vec{\theta}$. We then apply three successive transformations. The first transformation accounts for the size of the PSF, yielding the photon position $\vec{\theta}'$. The second transformation, which maps $\vec{\theta}'$ to $\vec{\theta}''$, induces ellipticity, skewness, triangularity and kurtosis into the simulated light distribution. The last transformation from $\vec{\theta}''$ to $\vec{\theta}'''$ accounts for shifts of the center of the profile. The coordinates $\vec{\theta}'''$ are the final position of the photon on the simulated image.

\subsection{Base profile selection}

In this section, we explain how we choose the base profile of our model. As detailed above, we use a fixed mixture of two Moffat profiles parameterized by three parameters $\beta_1, \beta_2, \gamma$. Since our base profile is intended to account for the first component in figure \ref{fig:svd-sdss}, we fit it to this mean PSF image from SDSS. To obtain a good fit, we also vary the FWHM of the base profile during the fitting procedure, but we do not use the resulting value later on. We perform the fit with the Particle Swarm Optimizer included in the \texttt{CosmoHammer} package \cite{Akeret2013} and find the best-fit values of $\beta_1 = 1.233$, $\beta_2 = 3.419$, $\gamma = 0.387$.

In the remainder of the paper, we use the base profile defined by these parameter values. Since the mean light distribution we compute from our sample is not perfectly symmetric, we do not obtain a perfect fit. However, this is not an issue here, since the distortion operations we apply after sampling from the base profile account for deviations from circular symmetry.

\section{Parameter estimation with a Convolutional Neural Network}
\label{sec:cnn}

In this section, we describe how we estimate the free parameters of our model to reproduce a given star. This problem is the inverse of the forward-modeling process introduced in section \ref{sec:psf-model}, namely how can we determine the best possible parameter values in order to model a given star? To solve this issue, we use a supervised machine learning technique, specifically a CNN. In the following, we first give a short introduction to Deep Learning and neural networks. We then go into detail about the specific network we use in this paper and explain the training strategy used to learn the parameters of the neural network.

\subsection{Deep Learning and Convolutional Neural Networks}

In this paper, we focus on supervised machine learning, which requires a set of input examples -- images in our case -- as well as the desired outputs, such as regression parameters. Given these input/output pairs, a machine learning model is trained to reproduce the output corresponding to each specific input in our dataset. In practice, many choices are left to the user such as the choice of model or the metric function used to evaluate its performance. We will here give a brief overview and refer the reader to \cite{Murphy2012, LeCun2015, Goodfellow2016} for a detailed survey of the relevant literature.

One specific type of model that we consider in this work is a deep neural network, which is part of a broader family of machine learning methods based on learning feature representations from data. Unlike other models, neural networks do not require explicitly constructing features from the input data but can directly learn patterns from the raw input signal. In a nutshell, this is accomplished by using a complex combination of neurons organized in nested layers that can learn a non-linear function of the input data.

\begin{figure}[tbp]
\centering
\includegraphics[width=\textwidth]{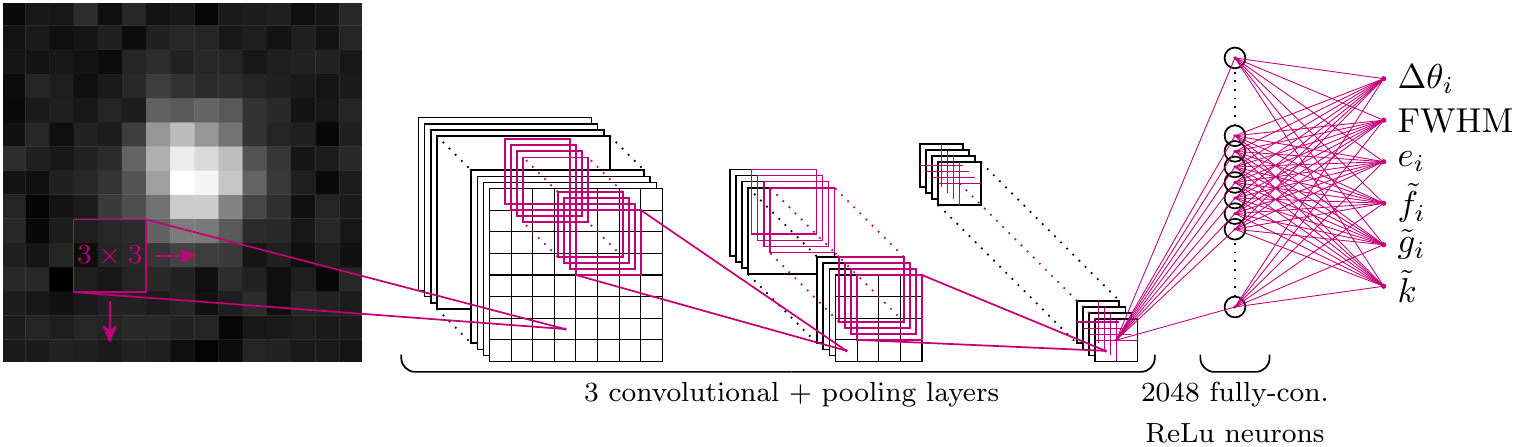}
\caption{Sketch of the CNN architecture used in this paper. The network consists of multiple convolutional and pooling layers that produce a rich set of features. These are passed to a fully-connected layer, which computes the final output of the CNN. See section \ref{subsec:implementation-architecture} for a detailed description of the architecture and section \ref{subsec:training-strategy} for details on the output of the CNN.}
\label{fig:cnn}
\end{figure}

In this work, we use a special type of neural network known as Convolutional Neural Network (CNN), see e.g. \cite{Rawat2017, Liu2017, McCann2017}, which has recently been achieving state-of-the-art performance on a variety of pattern-recognition tasks. One task where CNNs have particularly excelled is image classification \cite{Krizhevsky2012}, achieving results close to human accuracies \cite{Simonyan2014}. Three main types of layers are used to build a CNN architecture: convolutional layers, pooling layers and fully connected layers. Additional non-linear activation layers are used to construct non-linear models. Figure \ref{fig:cnn} shows a schematic representation of the CNN we use in this paper, which has a standard architecture. In the following, we give a brief description of the different layers of a CNN and their functions. We refer the reader to \cite{Goodfellow2016} for further details.

Convolutional layers perform discrete convolution operations on their input. These layers are parametrized by a set of learnable filters whose width and height are tunable parameters (we use $3 \times 3$ filters as detailed in the next section) and whose depth extends to the same depth as the input. When feeding the image to the first convolutional layer in the neural network, each filter is slid across the image and for each spatial position, an output value is produced by computing the correlation between the filter and the image content. This process yields one output map for each filter, each having the same size as the input image (assuming that we do not crop the boundaries which might require duplicating boundary pixels). During training, the network learns filters that will activate the most when seeing certain image properties that can discriminate the signal. As each layer has access to a set of several filters, the network will specialize each filter to recognize different image characteristics.

The set of outputs maps produced by each layer is then passed on to the next layer. The layer following a convolutional layer is typically a pooling layer that performs a downsampling operation on the input by merging local input patches, e.g. using an average or maximum operator. For example, a standard pooling layer would subdivide the input map into blocks and take the maximum value in each block, therefore reducing the size of the map as well as the amount of parameters (and computation) in the network. As shown in figure \ref{fig:cnn}, convolutional and pooling layers are repeated in a sequential fashion multiples times in order to construct a rich set of features from the input signal. Non-linear activation functions are used in between layers to enable the model to learn a non-linear function of the input. The resulting set of output maps (aka features) extracted by this sequence of layers is then passed on to a fully connected layer, which performs the high-level reasoning in order to produce the final output. This operation can typically be computed as a matrix multiplication. If used as the last layer in the network, it outputs a set of scores, one for each output parameter. In the next sections, we will detail the specific architecture used for our experiments as well as the training procedure used to learn the parameters of the neural network.

\subsection{\label{subsec:implementation-architecture}Implementation and network architecture}

The architecture used for our experiments is presented in figure \ref{fig:cnn}. It includes three consecutive convolutional layers, whereby each layer uses square filters with a side length of three pixels. The first convolutional layer directly processes the input image using 32 filters. Each $3 \times 3$ filter is applied to the input image using a convolution operation as explained in the previous section. We then repeat this operation in a sequential fashion, whereby we double the number of filters for each following layer, thus we use 64 and 128 filters in the second and third layer. Furthermore, each convolutional layer has adjustable additive bias variables, one per filter. The output of each convolutional layer is passed to a rectified linear unit (ReLU, \cite{Glorot2011}) followed by a pooling layer. As mentioned earlier, each pooling layer performs a max operation over sub-regions of the extracted feature maps resulting in downsampling by a factor of two. After the three stages of convolutions, activations and pooling, we place a fully connected layer with 2048 ReLU neurons. This step maps 128 feature maps with dimensions of $2 \times 2$ pixels (the outputs of the third convolutional and pooling layer) to 2048 features. The fully connected layer also includes an additive bias term with 2048 adjustable variables. Finally, these 2048 features are mapped to our 10-dimensional output space using a final matrix multiplication and bias addition. The CNN described in this section has approximately $1.16 \cdot 10^6$ trainable parameters. We use the \textsc{TensorFlow} package\footnote{Useful tutorials for \textsc{TensorFlow} can be found here: \url{https://www.tensorflow.org/tutorials/}. One tutorial particularly useful for this paper can be accessed via \url{https://www.tensorflow.org/tutorials/layers} (accessed on 22/02/2018).} \cite{Abadi2016} to implement our CNN.

\subsection{\label{subsec:training-strategy}Training strategy}

To learn the parameters $\theta$ of the neural network, we measure its performance using a loss function $L(\theta)$. The optimal network parameters $\theta_\text{opt}$ are found by minimizing $L$ with respect to $\theta$, i.e. $\theta_\text{opt} = \min_\theta L(\theta)$. The loss function we use is the Huber loss \cite{Huber1964}, defined as
\begin{align}
L_i(\theta) &= \begin{cases} \frac{1}{2} \left| y^\text{e}_i(\theta) - y^\text{t}_i\right|^2, & \left| y^\text{e}_i(\theta) - y^\text{t}_i \right| \leq \delta, \\
                                          \delta \left| y^\text{e}_i(\theta) - y^\text{t}_i \right| - \frac{1}{2} \delta^2, & \left| y^\text{e}_i(\theta) - y^\text{t}_i \right| > \delta,
                       \end{cases}  \\
L(\theta) &= \frac{1}{N_\text{dim}} \sum^{N_\text{dim}}_{i=1} L_i(\theta). \label{eq:huber-loss}
\end{align}
$y^\text{e}(\theta)$ is a vector of input parameters for our PSF model estimated by the network and $y^\text{t}$ is the corresponding vector of true values. $i$ indexes the dimensions $N_\text{dim} = 10$ of the parameter space of our PSF model and $|\cdot|$ denotes the absolute value. We use this loss function because it is more robust to outliers than the mean squared error (MSE). It is also differentiable everywhere due to the transition to the MSE at residual values smaller than $\delta = 10^{-5}$. When we started experimenting with the CNN, we first used the MSE as loss function. However, we found that this choice led to biases in the parameters estimates produced by the network. We therefore switched to a more robust loss function, which reduced the biases, confirming that they were caused by outliers.

In order to optimize the Huber loss defined in eq. \eqref{eq:huber-loss}, we use a variant of stochastic gradient descent (e.g. \cite{Ruder2016}) called \textsc{Adam} \cite{Kingma2014}. We set the learning rate $\alpha = 0.001$ and the other hyper-parameters of the \textsc{Adam} optimizer to $\beta_1 = 0.9$, $\beta_2 = 0.999$ and $\epsilon = 10^{-8}$, as recommended by \cite{Kingma2014}. The gradients of the loss function are computed using the back-propagation algorithm implemented in \textsc{TensorFlow} \cite{Abadi2016}.

Furthermore, we train our network to learn a rescaled version of the parameters we aim at estimating to ensure that each dimension of the parameter space contributes approximately equally to the loss function. This is accomplished by rescaling all parameters to the same numerical range, i.e. $y^\text{t}_i \in [-1, 1]$.

To simplify the mapping our CNN has to learn, we apply certain transformations to parts of the parameter space. As can be seen from eq. \eqref{eq:transformation-derivatives}, $f_1$, $f_2$, $g_1$, $g_2$ have units of $\text{pixel}^{-1}$ and $k$ has units of $\text{pixel}^{-2}$. This means that the change in shape induced by the corresponding transformations depends on the size of the simulated star. In other words, the parameters $f_1$, $f_2$, $g_1$, $g_2$, $k$ are correlated with the size of the simulated PSF. In order to remove this correlation as much as possible, we train our network to learn dimensionless versions of the parameters, which we denote with $\tilde{f}_1, \tilde{f}_2, \tilde{g}_1, \tilde{g}_2, \tilde{k}$. They are obtained via the following transformations:
\begin{align}
\tilde{f}_1, \tilde{f}_2, \tilde{g}_1, \tilde{g}_2 &= \left\{ f_1, f_2, g_1, g_2 \right\} \cdot \text{FWHM}, \\
\tilde{k}&= k \cdot \text{FWHM}^2.
\end{align}

\subsection{\label{subsec:training-data}Training data}

The data used to train the network is obtained by uniformly sampling a hypercube in our parameter space and simulating a star for each sample. The ranges of the parameters defining the hypercube are denoted in table \ref{tab:ranges-training-cube}. The training data we generate has the same format as our sample of stars from SDSS, meaning that we simulate our training stars using a $15 \times 15$ pixel grid. Initially, each star is placed in the center of the grid (before applying the offsets $\Delta \theta_i$).

\begin{table}[htb]
\centering
\begin{tabular}{| c | c |}
\hline
 & Range \\
\hline
$\Delta \theta_i$ & $-0.5, 0.5$ pixel\\
FWHM & $1.5, 6.0$ pixel \\
$e_i$ & $-0.25, 0.25$ \\
$\tilde{f}_i$ & $-0.25, 0.25$ \\
$\tilde{g}_i$ & $-0.1, 0.1$ \\
$\tilde{k}$ & $-0.4, 0.4$ \\
\hline
\end{tabular}
\caption{\label{tab:ranges-training-cube}Range of each parameter defining the volume in parameter space that we sample uniformly to create the training data for our CNN.}
\end{table}

Given a sample of parameter values, we additionally need to specify the number of photons used to simulate the star. To this end, we also draw random magnitudes between 15 and 20.5, random gain values between 4.5 and 5$\, e^- / \text{ADU}$ and random magnitude zeropoints between 28 and 28.5. We adopt these ranges because they include the majority of stars in our SDSS sample, such that our training data is representative of the SDSS data. When calculating the number of photons from the magnitude, the gain and the magnitude zeropoint, we include Poisson noise.

We draw $10^8$ samples from the hypercube defined above and evaluate our PSF model for each sample in order to obtain a training dataset. Since real images contain background noise, we have to account for this component in order to obtain a network that can be reliably applied to survey data. We add Gaussian noise with mean zero to our simulated stars on the fly when feeding them to network. We sample the corresponding standard deviations uniformly between 3.5 and 5.5$\,$ADUs, since the background level of most of the SDSS images we use is located in this interval. After adding noise, we normalize the simulated images to have a maximum intensity of 1, as was done for the real data (see section \ref{sec:sdss-star-sample}).

As mentioned in section \ref{subsec:training-strategy}, we rescale the parameters the network is supposed to estimate before training the CNN. We re-center the samples drawn from the hypercube to have zero mean and adjust the numerical range to extend from -1 to 1 along each dimension. We then train our CNN by feeding it with batches of simulated stars and the corresponding rescaled true parameters, whereby we set the batch size to 2000 stars. We iterate through three training epochs, meaning that each training star is given to the network three times, each time with a different background noise level. Thus, the network is trained on $3 \cdot 10^8$ stars in total, which takes around $1.5$ days on a machine with 24 CPU cores (no GPUs were used in this work).

\subsection{\label{subsec:cnn-output-corrections}CNN output corrections}

We use the \textsc{Adam} optimizer to minimize the Huber loss for each batch of training stars given to the network. This is accomplished by adjusting the weights of the CNN to reduce both the variance with which the estimated values scatter around the true values as well as the biases in the estimated values. These two objectives might not be completely compatible, i.e. there can be a bias-variance trade-off. For example, the optimizer might reduce the spread in one parameter, but the corresponding adjustments in the weights of the CNN lead to a small increment in the bias of this parameter. To correct for this effect, we use linear fits that map the estimated, potentially slightly biased parameter values closer to the true values. These fits are performed after the training is complete. When applying our CNN to estimate the parameters for our PSF model, we always correct the estimates using these linear fits. To obtain the linear corrections, we use $50\,000$ randomly selected stars from the training sample. We perform the fits with the random sample consensus (RANSAC, \cite{Fischler1981}) method, which is robust to outliers. RANSAC fitting is implemented for example within the \textsc{scikit-learn} library \cite{Pedregosa2011}, which we use for this purpose.

We originally introduced these linear corrections to compensate for biases in the parameter values estimated by the network caused by outliers when training with the MSE instead of the Huber loss (see section \ref{subsec:training-strategy}). After switching to the Huber loss, these biases were reduced. Accordingly, the slopes and intercepts found by the RANSAC algorithm are now very close to one and zero, respectively. Thus, the linear corrections change the parameters predicted by the CNN on the percent to sub-percent level only.

\section{\label{sec:results}Results}

Here, we present the results from training our CNN introduced in section \ref{subsec:implementation-architecture} and using it in combination with the PSF model detailed in section \ref{subsec:model-description} to render a model image of each SDSS star in our sample. To train the network on our training data (see section \ref{subsec:training-data}), we follow the strategy explained in section \ref{subsec:training-strategy}. After the training is complete, we apply the CNN to our SDSS dataset. The resulting parameter estimates are fed into the PSF model to render a model image of each SDSS star. To compute the number of photons used to render a SDSS star, we use the measured magnitude of the object as well as the magnitude zeropoint and the gain of the corresponding exposure. 

We first present the performance of the network in conjunction with the PSF model on the SDSS data. We then give results concerning the validation of our network and compare our method to a direct fit of the PSF model to SDSS stars. Finally, we compare our PSF modeling approach to alternative approaches and conclude with an analysis of the impact of our modeling errors on cosmic shear measurements.

\subsection{\label{subsec:perfomance-sdss}Performance on SDSS}

\begin{figure}[tbp]
\centering
\includegraphics[width=\textwidth]{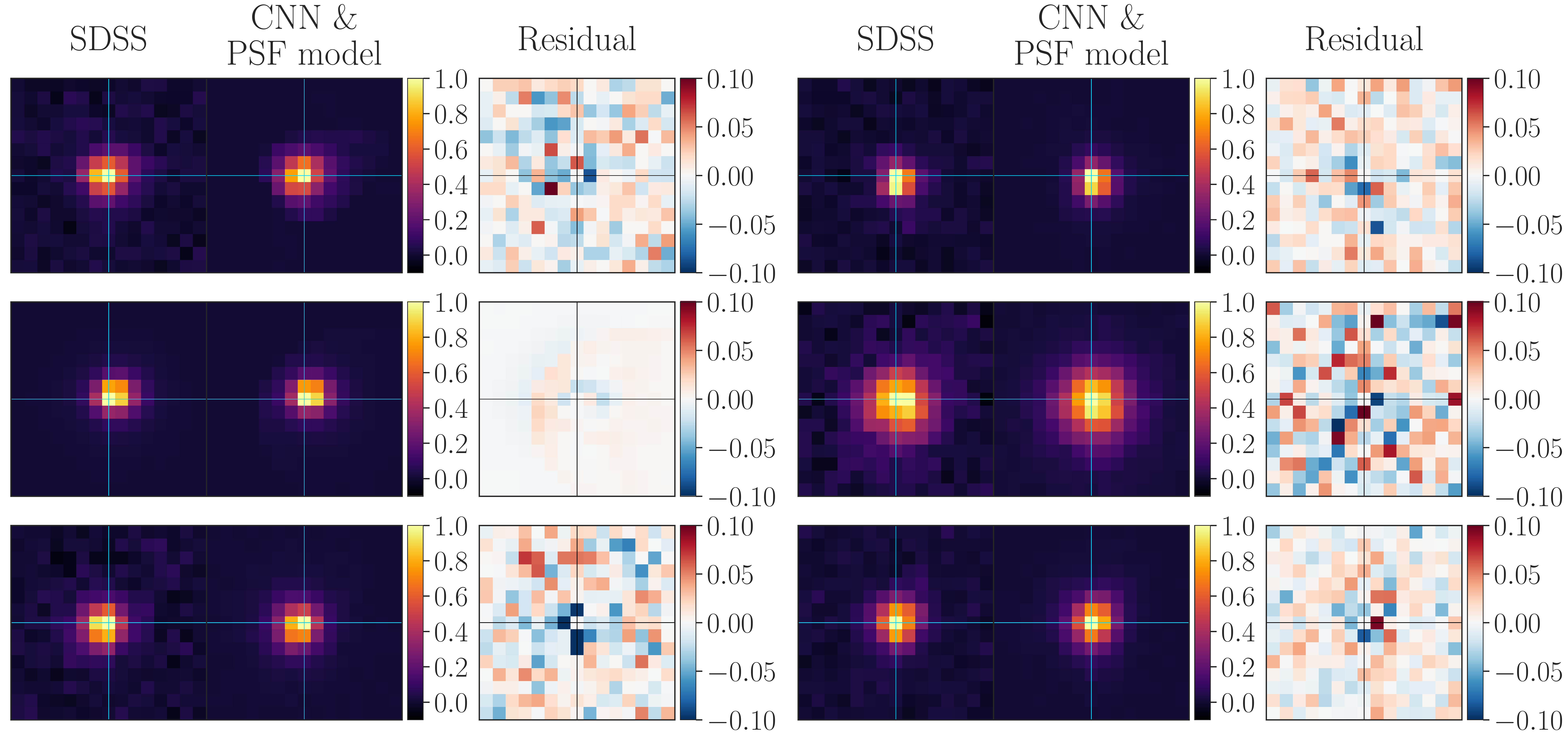}
\caption{\label{fig:sdss-stars--vs--cnn-stars}Six randomly selected stars from the SDSS sample compared to their model images. The left columns show the image cutouts of the objects extracted from real SDSS images. The model images are displayed in the middle columns. They are obtained by applying the trained network to the SDSS data and evaluating our PSF model using the resulting parameter estimates. Both the SDSS as well as the modeled images have been normalized to a maximum pixel intensity of 1. The right columns show the residuals that remain after subtracting the SDSS images from the model images.}
\end{figure}

In figure \ref{fig:sdss-stars--vs--cnn-stars}, we compare six randomly selected stars from our SDSS sample to their model images. We see that our model is able to describe a wide range of PSF images with diverse sizes and shapes. This can also be seen from the residual images, which mainly consist of photon and background noise.

\begin{figure}[tbp]
\centering
\includegraphics[width=\textwidth]{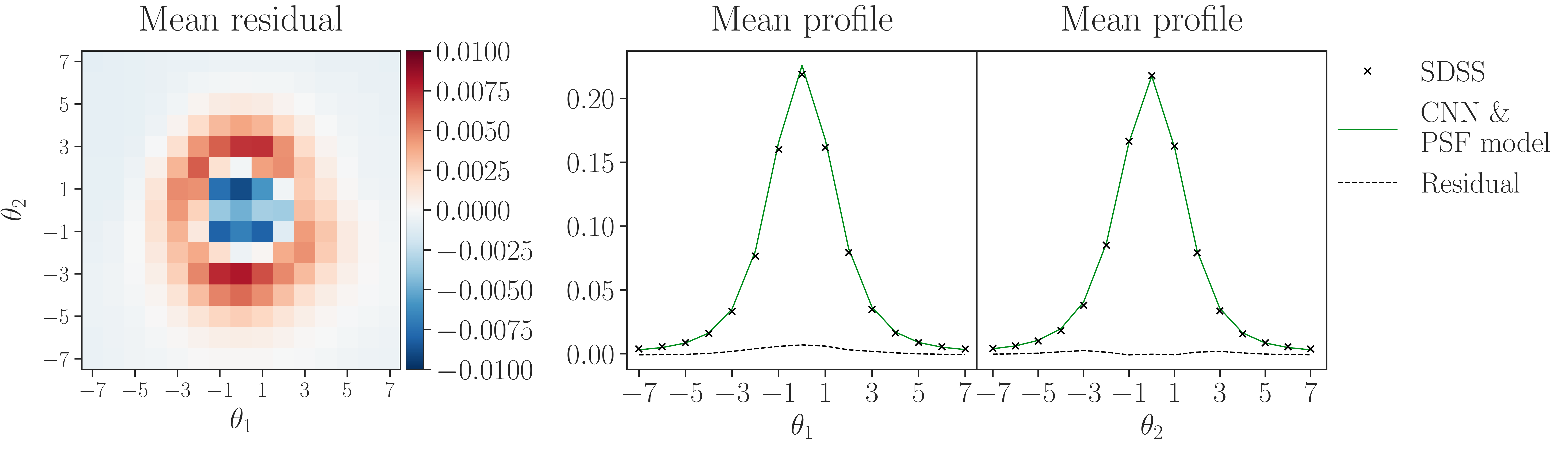}
\caption{\label{fig:mean-residual}Mean residual fluxes after subtracting the SDSS stars from their simulated counterparts. The left panel shows the mean residual star obtained from averaging the pixel values of the individual residuals. The middle and the right panel show the mean profiles of the SDSS and the simulated data, which are obtained by averaging the pixel values in $\theta_1$- and $\theta_2$-direction.}
\end{figure}

To give a global overview of the performance of our model in conjunction with the network, we show the average residual image in figure \ref{fig:mean-residual}. As can be seen, there is structure beyond photon and pixel noise left in the mean residual star. However, the average residuals peak at the percent level compared to the peak of the actual image. Furthermore, the projected profiles in the $\theta_1$- and $\theta_2$-direction (see the right panels of figure \ref{fig:mean-residual}) show that the SDSS stars are well reproduced by our model. Additionally, we find that the principal components constructed from our modeled stars are consistent with those from the SDSS data (see appendix \ref{app:svd-cnn}), further validating our method.

\subsection{\label{subsec:validation}Validation of the network}

To test further the performance of the CNN, we apply it to simulated data for which we know the ground truth model parameters. Importantly, the network has never seen this validation data during the training, such that it did not have the chance to accidentally overfit and adjust specifically to this dataset. We use two types of validation sets: one training-like validation set and one SDSS-like validation set. The training-like validation sample is statistically identical to the training data and we use it to show that our network did not overfit during the training phase. The SDSS-like validation set on the other hand is statistically close to real survey data. Thus, this validation sample allows us to probe the performance of the CNN in the regime we are interested in.

To generate the training-like validation set, we draw $200\,000$ samples from the hypercube defined in table \ref{tab:ranges-training-cube} and evaluate our PSF model for these samples. Also the photon and background noise properties of this validation sample are the same as for the training data. The SDSS-like validation sample is created from the modeled SDSS stars described above in section \ref{subsec:perfomance-sdss}. We add Gaussian background noise that corresponds to standard deviations uniformly distributed between 3.5 and 5.5$\,$ADUs (the same interval was used for training the network, see section \ref{subsec:training-data}). We then apply the network to both validation samples and obtain two sets of estimated model parameters than can be compared to the input parameters used to generate the validation data.

\begin{figure}[tbp]
\centering
\includegraphics[width=\textwidth]{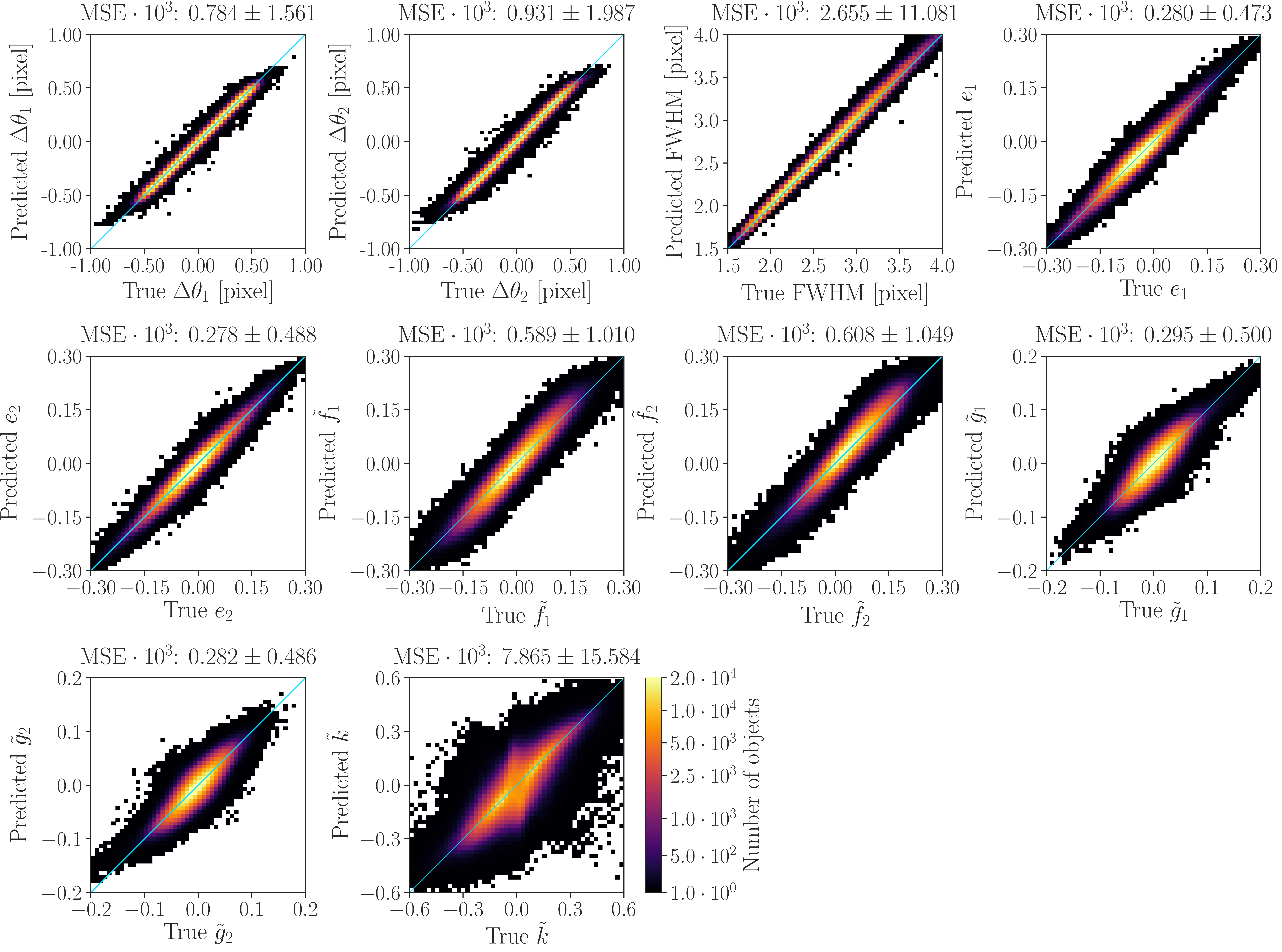}
\caption{\label{fig:true-vs-pred}Validation of the performance of the trained CNN. To produce this figure, we apply the network to simulated SDSS-like stars, which are described above in the text. For each object, we then have two sets of parameters available: the ground truth input values used to simulate the star and the values estimated by the CNN. We plot the input values on the $x$- and the estimated values on the $y$-axis. The blue diagonal lines correspond to an ideal one-to-one relation between input and measured values. The displayed color scale applies to all panels. We furthermore report the MSE of our network for each parameter. The given uncertainties correspond to one standard deviation.}
\end{figure}

In figure \ref{fig:true-vs-pred}, we compare the model parameters estimated from the SDSS-like validation data to the input parameters used to generate this validation set. For most of the objects, the estimates scatter tightly around the blue diagonal lines that represent the ideal one-to-one relation. This means that the network is able to measure reliable parameter values for the majority of stars in our SDSS-like validation sample. Furthermore, we note that the CNN performs very well for the lowest-order distortions, which are size (FWHM) and ellipticity ($\tilde{e}_i$). This is especially important in the context of weak lensing measurements, since the shear signal is very prone to biases in these parameters. Finally, it can be seen from figure \ref{fig:true-vs-pred} that the parameter on which the CNN performs less well is the kurtosis $\tilde{k}$. This can be directly linked to the mean residual shown in figure \ref{fig:mean-residual}, which displays the typical signature  caused by small errors in the steepness of the light distribution.

\begin{figure}[tbp]
\centering
\includegraphics[width=0.6\textwidth]{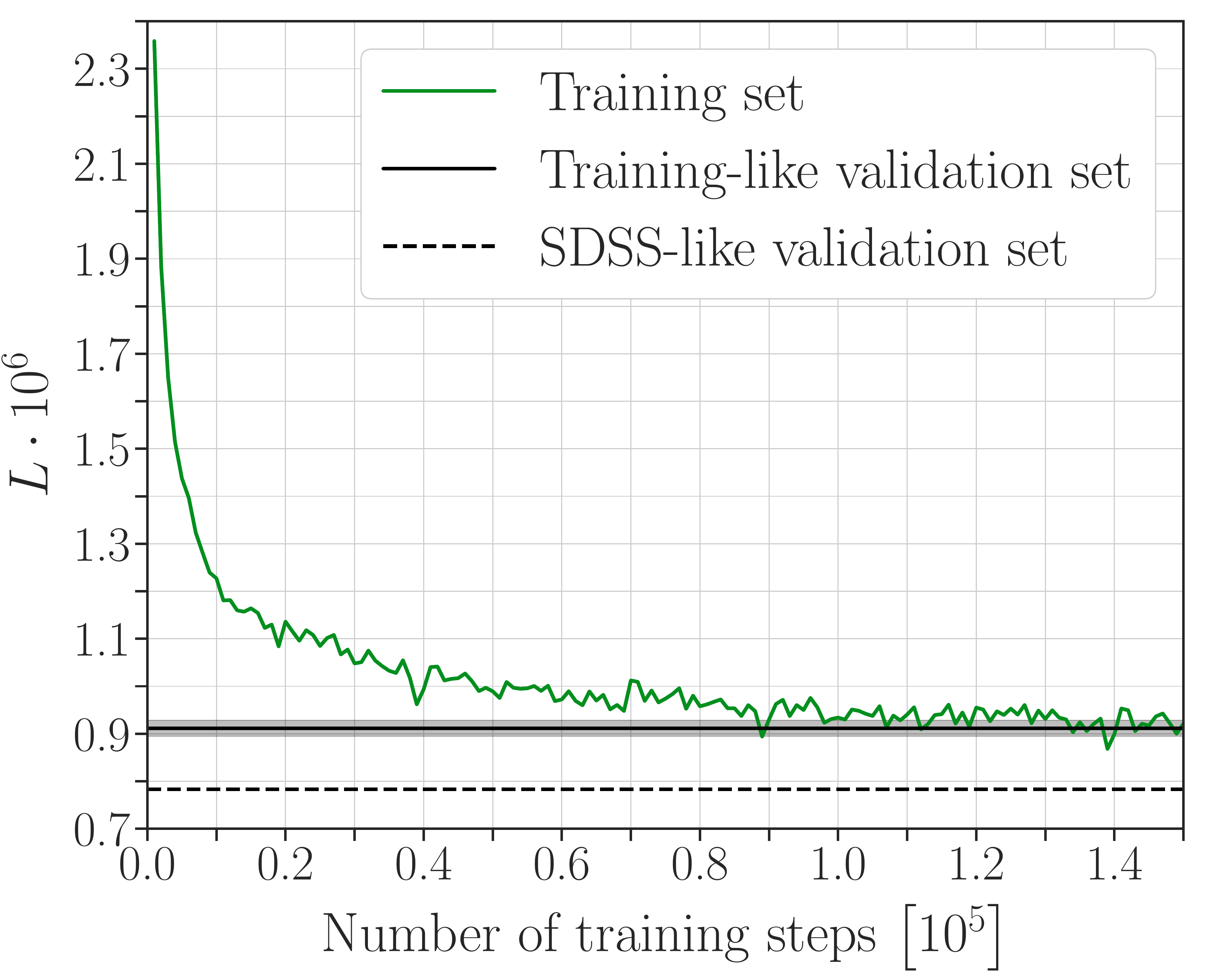}
\caption{\label{fig:loss}Loss function $L$ used to train the CNN (see eq. \eqref{eq:huber-loss}) as a function of the number of training steps (green line). Each data point corresponds to the average loss of one batch of $2\,000$ training stars. We furthermore show the loss function evaluated on the two validation samples described in this section (black lines). To compute the loss of the training-like validation set, we first split this sample into 100 batches of $2\,000$ stars, compute the per-batch losses and then average these values. This allows us to also compute an uncertainty on the loss of the training-like validation set in terms of one standard deviation of the 100 per-batch loss values, which is shown as the grey band. Note that for this figure, we did not apply the linear corrections explained in section \ref{subsec:cnn-output-corrections} to the parameter values estimated by the CNN from the two validation sets. We avoided the corrections for this particular figure in order to compare the training and the validation losses on an equal footing, since the corrections were not applied during the training phase either, see section \ref{subsec:cnn-output-corrections}.}
\end{figure}

In figure \ref{fig:loss}, we show the training error of the network (green line). Each data point corresponds to the average loss of one batch of $2\,000$ training stars (see section \ref{subsec:training-data}). As expected, the value of the loss function decreases as the number of processed training examples increases. The figure also shows the loss function evaluated on the two validation sets described above (black lines). To compute the loss of the SDSS-like validation set, we directly average the losses computed for the individual stars in this sample. To compute the loss of the training-like validation set, we first split this sample into 100 batches of $2\,000$ stars, which corresponds to the training batch size. We then compute the average loss of each individual batch and average these 100 loss values to obtain the value displayed in figure \ref{fig:loss}. This allows us to also compute an uncertainty on the loss of the training-like validation set, which is given by the grey band in the figure. It corresponds to one standard deviation of the 100 per-batch loss values computed from the split training-like validation set.

As can be seen from figure \ref{fig:loss}, the final training loss of the network is on the same level as the loss of the training-like validation sample. This means that our network did not overfit. We attribute this to the fact that we only iterated through four training epochs, such that the CNN saw each training star only four times. Furthermore, even though the network is trained on the same objects multiple times, the background noise is different each time (since it is added on the fly during the training phase). This effectively acts as a further regularization which prevents overfitting. Figure \ref{fig:loss} also shows that the loss of the SDSS-like validation sample is smaller than the training loss. This is because the SDSS-like validation set is statistically different from the training data. As can be seen from figure \ref{fig:true-vs-pred}, the SDSS-like validation stars cluster around zero for the ellipticity, flexion and kurtosis parameters. This region contains the highest density of training examples, since the training data is also centered on zero for the parameters $\tilde{e}_i$, $\tilde{f}_i$, $\tilde{g}_i$ and $\tilde{k}$. Thus, most of the SDSS-like validation stars lie in the regions where the network has been trained most extensively and only few objects are close to the boundaries of the hypercube the training data was sampled from. Contrary to this, the fraction of stars close the boundaries of the hypercube is much higher for the uniformly distributed training data. This results in the smaller loss value of the SDSS-like validation set seen in figure \ref{fig:loss}.

\subsection{\label{subsec:comparison-direct-fitting}Comparison with direct fitting}

In this section, we briefly compare the performance of our CNN to a direct fitting approach, which could be an alternative to using a neural network. We use the same fixed base profile as for the CNN, but instead of using the network to estimate the ten free parameters of our PSF model, we now use an optimization algorithm. The cost function we minimize is the sum of residual pixel values (MSE). To perform the fits, we use the function \texttt{scipy.optimize.minimize}, which is part of the SciPy library \cite{Jones2001}. We try out all applicable optimization algorithms that are available for \texttt{scipy.optimize.minimize} and report the results obtained with the algorithm that yields the best average cost function. Our findings are detailed in appendix \ref{app:direct-fitting}. In a nutshell, we find that direct fitting yields results similar to the ones obtained with the CNN in terms of residual pixel values. However, the big disadvantage of direct fitting is the runtime. We found that it takes on average several seconds to fit one star, whereby the runtime varies heavily from star to star and can also be significantly larger. Thus, applying this method to large wide-field datasets with hundreds of thousands to millions of stars would be quite intense in terms of computing time. Furthermore, this runtime issue renders direct fitting practically intractable in the context of forward modelling and $\textit{MCCL}$, where one typically has to simulate and analyze data volumes even larger than the actual survey data volume. The CNN on the other hand can process thousands of stars in one second, making this method ideal for forward modeling applications.

\subsection{Comparison with other PSF modeling techniques}

To place our results in a global context, we compare them to results obtained with other PSF modeling techniques. \cite{Annis2014} present coadded SDSS images and model their PSF. In this case, the PSF modeling is based on a Karhunen–Lo{\'e}ve expansion (cf. PCA). Figure 6 in \cite{Annis2014} shows the square root of the ratio of the PSF size measured on the SDSS images to the size measured on the modeled PSF. Using the validation sample presented in section \ref{subsec:validation}, we compute the mean of the square root of the ratio of the true PSF FWHM to the FWHM estimated by the network and find 0.996. This compares well to the ratio given in \cite{Annis2014}, which varies approximately between 0.990 and 1.005 for the $r$-band. We conclude that in terms of the PSF size, our method performs similarly well as the Karhunen–Lo{\'e}ve expansion used in \cite{Annis2014}.

We also compare our method to the PSF modeling in latest DES weak lensing analysis \cite{Zuntz2017}, which relies on a pixel-based approach with \textsc{PSFEx} \cite{Bertin2011}. Figure 7 in \cite{Zuntz2017} gives an impression of the corresponding size and ellipticity residuals. Even though \cite{Zuntz2017} uses a different definition of size and ellipticity than this work, we can still compare at least the orders of magnitude of our results. Using the validation sample described in section \ref{subsec:validation}, we find a mean squared size error of $\langle \delta \, \text{FWHM}^2 \rangle = 0.0179''$  and mean ellipticity residuals of $\langle \delta \, e_1 \rangle = 0.0005$ and $\langle \delta \, e_2 \rangle = 0.0001$. Since the residuals in the DES analysis are of the same order of magnitude, we conclude that in terms of these numbers, our method performs similarly well as \textsc{PSFEx} on DES data.

\subsection{Approximate cosmic shear forecast}

Here, we examine the impact of our modeling inaccuracies onto weak lensing measurements. To this end, we calculate the systematic uncertainty $\sigma_\text{sys}^2$ on a shear signal resulting from our PSF modeling errors. To compute this quantity, we use equation (15) from \cite{Paulin-Henriksson2008}. There are two simplifications related to using this equation: (i) \cite{Paulin-Henriksson2008} assume that an unbiased estimator of size and ellipticity is available, which is of course not exactly true for our CNN, (ii) the size and ellipticity in \cite{Paulin-Henriksson2008} correspond to the second-order moments of the PSF light distribution. Due to the presence of flexion and kurtosis, the size and the ellipticity estimated by our network correspond only approximately to second-order moments. However, since we are only aiming at an order-of-magnitude forecast, we are confident that equation (15) from \cite{Paulin-Henriksson2008} can still give us an idea of the performance of our estimator.

To compute $\sigma_\text{sys}^2$, we use $P^\gamma = 1.84$ and $\langle \left| e_\text{gal} \right|^2 \rangle = 0.16$, as recommended by \cite{Paulin-Henriksson2008}. We furthermore set $R_\text{gal} / R_\text{PSF} = 1.5$, since one typically uses only galaxies sufficiently larger than the PSF for shear measurements. For the mean PSF size and ellipticity, we use the mean values estimated on the SDSS data. Finally, we use the SDSS-like validation data described in section \ref{subsec:validation} to estimate the variance in the error of our predicted sizes and ellipticities. This yields a systematic uncertainty of $\sigma_\text{sys}^2 = 3.25 \cdot 10^{-5}$. This number can be compared to a requirement on $\sigma_\text{sys}^2$ for a SDSS-like survey to not be systematics-limited. In \cite{Nicola2016}, SDSS coadded data was used in combination with other datasets to constrain cosmology. The corresponding survey area is $A_\text{s} = 275\,\text{deg}^2$, the surface density of lensing galaxies is $n_\text{g} \approx 3.2\,\text{arcmin}^{-2}$, the median redshift of the lensing sample is $z_\text{m} \approx 0.6$ and the maximum multipole used for the power spectra is $\ell_\text{max} = 610$. Following \cite{Amara2007}, we translate these numbers into a requirement onto the systematics and obtain $\sigma_\text{sys}^2 \lesssim 1.12 \cdot 10^{-5}$.

This number is slightly smaller than the systematic uncertainty computed above, however, both numbers are of the same order of magnitude. Furthermore, it is important to notice that we model the PSF of individual SDSS exposures in this work. Contrary to that, the requirements we computed correspond to SDSS coadded data, which generally has a better PSF quality than the individual exposures. Also, one would typically average over several stars in the same region to estimate the PSF at a given position, which would further reduce the systematic uncertainty induced by our modelling inaccuracies. We therefore conclude that the performance of our method is on a level suitable for weak-lensing applications.

The requirements for state-of-the-art weak lensing surveys such as DES and KiDs are significantly lower than the numbers computed above. For a DES Y1-like survey, we have $\sigma_\text{sys}^2 \lesssim 3 \cdot 10^{-6}$, see \cite{Bruderer2017}. For KiDS-450, we find $\sigma_\text{sys}^2 \lesssim 5 \cdot 10^{-6}$. To compute this number, we use $A_\text{s} = 360.3\,\text{deg}^2$, $n_\text{g} = 8.53\,\text{arcmin}^{-2}$, $z_\text{m} \approx 0.6$ and $\ell_\text{max} = 1000$. We took these values from \cite{Hildebrandt2017}, whereby we compute $z_\text{m}$ as the mean of the median redshifts in the four tomographic bins of the KiDS analysis. For both DES and KiDS, $\sigma_\text{sys}^2$ is one order of magnitude smaller than the value we computed for our model ($\sigma_\text{sys}^2 = 3.25 \cdot 10^{-5}$). However, both surveys have a better resolved PSF than SDSS, such that one would expect the CNN to perform better on data from DES or KiDS. Furthermore, as mentioned above, one typically averages over multiple stars to estimate the PSF. Both DES and KiDS are deeper than SDSS, such that the density of stars used for PSF estimation is higher. Finally, both surveys take multiple exposures per field of view, such that multiple samples of the PSF at each position are available (or alternatively coadded data). This increases the quality of the PSF estimation further. For these reasons, we are confident that our method is suitable for state-of-the-art cosmological applications.

\section{\label{sec:conclusion}Conclusion}

We presented a novel approach to model the PSF of wide-field surveys. Our method relies on a data-driven PSF model, which is able to describe a wide range of PSF shapes and on Deep Learning to estimate the free parameters of the model. The combination of these two approaches allows us to render the PSF of wide-field surveys at the pixel level. This can be used, for instance, within the \textit{MCCL} framework to generate realistic wide-field image simulations, which are useful for calibrating shear measurements.

We demonstrated our method with publicly available SDSS data. To construct our model, we performed a PCA on our SDSS sample of stars. This allowed us to incorporate the main modes of variation extracted from the data into our model. To this end, we implemented the model in two steps. First, we chose a fixed base profile that is kept constant for all stars. Then, we implemented a set of perturbative distortion operations that account for variations in the PSF. These distortions were designed to enable our model to reproduce the most important principal components extracted from the SDSS sample.

To estimate the free parameters of our PSF model from survey data, we opted for a CNN. This allows for very fast measurements once training is complete, which is vital for incorporating our method into the \textit{MCCL} scheme. We trained the network on a large number of simulated stars and subsequently applied it to the SDSS data. We then generated a model image of each star in our sample. By comparing the modeled star images to their counterparts from SDSS, we demonstrated that our model in conjunction with the CNN is able to reliably reproduce the SDSS PSF. Furthermore, using the simulated version of the SDSS sample, we validated our network and demonstrated that it is able to produce largely unbiased parameter estimates for the big majority of objects in our sample. This was ensured by applying linear corrections to the estimates made by the network. The correction factors were found using RANSAC fits after training the network.

The results presented in this paper offer encouraging prospects for applying our approach to state-of-the-art imaging surveys. Since currently ongoing survey programs have in general a better resolved PSF than SDSS, we expect our approach to perform at least as well as it did on the dataset used here. Future improvements include training the neural network on larger amounts of training data, possibly using GPUs (in this work we only used CPUs for training). Furthermore, if necessary, our model can be extended to include additional higher-order distortions.

\acknowledgments

This work has made use of data from the European Space Agency (ESA) mission \textit{Gaia} (\url{https://www.cosmos.esa.int/gaia}), processed by the \textit{Gaia} Data Processing and Analysis Consortium (DPAC, \url{https://www.cosmos.esa.int/web/gaia/dpac/consortium}). Funding for the DPAC has been provided by national institutions, in particular the institutions participating in the \textit{Gaia} Multilateral Agreement. We acknowledge support by grant number 200021\_169130 from the Swiss National Science Foundation. \\
This research made use of the Python packages \textsc{NumPy} \cite{vanderWalt2011}, \textsc{h5py}\footnote{\url{http://www.h5py.org}}, \textsc{Matplotlib} \cite{Hunter2007}, \textsc{seaborn} \cite{Waskom2017}, \textsc{jupyter} \cite{Kluyver2016}, \textsc{ipython} \cite{Perez2007}, \textsc{Astropy} \cite{AstropyCollaboration2018}, \textsc{Numba} \cite{Lam2015} and \textsc{pandas} \cite{McKinney2011}.

\bibliographystyle{JHEP}
\bibliography{psf_model_cnn}

\providecommand{\href}[2]{#2}\begingroup\raggedright\begin{thebibliography}{10}

\bibitem{Bradt2003}
H.~{Bradt}, \emph{{Astronomy Methods : A Physical Approach to Astronomical
  Observations}}.
\newblock Cambridge University Press, Cambridge, MA, USA, 2003,
  \href{http://dx.doi.org/{10.1017/CBO9780511802188
  }}{{10.1017/CBO9780511802188 }}.

\bibitem{Haeussler2007}
B.~{H{\"a}ussler}, D.~H. {McIntosh}, M.~{Barden}, E.~F. {Bell}, H.-W. {Rix},
  A.~{Borch} et~al., \emph{{GEMS: Galaxy Fitting Catalogs and Testing
  Parametric Galaxy Fitting Codes: GALFIT and GIM2D}},
  \href{http://dx.doi.org/{10.1086/518836}}{\emph{The Astrophysical Journal
  Supplement Series} {\bf 172} (October, 2007) 615--633},
  [\href{http://arxiv.org/abs/arXiv:0704.2601}{{\tt arXiv:0704.2601}}].

\bibitem{Gabor2009}
J.~{Gabor}, C.~{Impey}, K.~{Jahnke}, B.~{Simmons}, J.~{Trump}, A.~{Koekemoer}
  et~al., \emph{{Active Galactic Nucleus Host Galaxy Morphologies in COSMOS}},
  \href{http://dx.doi.org/{10.1088/0004-637X/691/1/705}}{\emph{The
  Astrophysical Journal} {\bf 691} (January, 2009) 705--722},
  [\href{http://arxiv.org/abs/arXiv:0809.0309}{{\tt arXiv:0809.0309}}].

\bibitem{Birrer2016}
S.~{Birrer}, A.~{Amara} and A.~{Refregier}, \emph{{The mass-sheet degeneracy
  and time-delay cosmography: analysis of the strong lens RXJ1131-1231}},
  \href{http://dx.doi.org/{10.1088/1475-7516/2016/08/020}}{\emph{Journal of
  Cosmology and Astroparticle Physics} {\bf 8} (August, 2016) 020},
  [\href{http://arxiv.org/abs/arXiv:1511.03662}{{\tt arXiv:1511.03662}}].

\bibitem{Wong2017}
K.~C. {Wong}, S.~H. {Suyu}, M.~W. {Auger}, V.~{Bonvin}, F.~{Courbin}, C.~D.
  {Fassnacht} et~al., \emph{{H0LiCOW - IV. Lens mass model of HE 0435-1223 and
  blind measurement of its time-delay distance for cosmology}},
  \href{http://dx.doi.org/{10.1093/mnras/stw3077}}{\emph{Monthly Notices of the
  Royal Astronomical Society} {\bf 465} (March, 2017) 4895--4913},
  [\href{http://arxiv.org/abs/arXiv:1607.01403}{{\tt arXiv:1607.01403}}].

\bibitem{Refregier2003b}
A.~{Refregier}, \emph{{Weak Gravitational Lensing by Large-Scale Structure}},
  \href{http://dx.doi.org/{10.1146/annurev.astro.41.111302.102207}}{\emph{Annual
  Review of Astronomy and Astrophysics} {\bf 41} (2003) 645--668},
  [\href{http://arxiv.org/abs/astro-ph/0307212}{{\tt astro-ph/0307212}}].

\bibitem{Hoekstra2008}
H.~{Hoekstra} and B.~{Jain}, \emph{{Weak Gravitational Lensing and Its
  Cosmological Applications}},
  \href{http://dx.doi.org/{10.1146/annurev.nucl.58.110707.171151}}{\emph{Annual
  Review of Nuclear and Particle Science} {\bf 58} (November, 2008) 99--123},
  [\href{http://arxiv.org/abs/arXiv:0805.0139}{{\tt arXiv:0805.0139}}].

\bibitem{Kilbinger2015}
M.~{Kilbinger}, \emph{{Cosmology with cosmic shear observations: a review}},
  \href{http://dx.doi.org/{10.1088/0034-4885/78/8/086901}}{\emph{Reports on
  Progress in Physics} {\bf 78} (July, 2015) 086901},
  [\href{http://arxiv.org/abs/arXiv:1411.0115}{{\tt arXiv:1411.0115}}].

\bibitem{Heymans2012}
C.~{Heymans}, B.~{Rowe}, H.~{Hoekstra}, L.~{Miller}, T.~{Erben}, T.~{Kitching}
  et~al., \emph{{The impact of high spatial frequency atmospheric distortions
  on weak-lensing measurements}},
  \href{http://dx.doi.org/{10.1111/j.1365-2966.2011.20312.x}}{\emph{Monthly
  Notices of the Royal Astronomical Society} {\bf 421} (March, 2012)
  {381--389}}, [\href{http://arxiv.org/abs/arXiv:1110.4913}{{\tt
  arXiv:1110.4913}}].

\bibitem{Paulin-Henriksson2008}
S.~{Paulin-Henriksson}, A.~{Amara}, L.~{Voigt}, A.~{Refregier} and S.~{Bridle},
  \emph{{Point spread function calibration requirements for dark energy from
  cosmic shear}},
  \href{http://dx.doi.org/{10.1051/0004-6361:20079150}}{\emph{Astronomy {\&}
  Astrophysics} {\bf 484} (June, 2008) {67--77}},
  [\href{http://arxiv.org/abs/arXiv:0711.4886}{{\tt arXiv:0711.4886}}].

\bibitem{Paulin-Henriksson2009}
S.~{Paulin-Henriksson}, A.~{Refregier} and A.~{Amara}, \emph{{Optimal point
  spread function modeling for weak lensing: complexity and sparsity}},
  \href{http://dx.doi.org/{10.1051/0004-6361/200811061}}{\emph{Astronomy {\&}
  Astrophysics} {\bf 500} (June, 2009) {647--655}},
  [\href{http://arxiv.org/abs/arXiv:0901.3557}{{\tt arXiv:0901.3557}}].

\bibitem{Amara2010}
A.~{Amara}, A.~{R{\'{e}}fr{\'{e}}gier} and S.~{Paulin-Henriksson},
  \emph{{Cosmic shear systematics: software-hardware balance}},
  \href{http://dx.doi.org/{10.1111/j.1365-2966.2010.16326.x}}{\emph{Monthly
  Notices of the Royal Astronomical Society} {\bf 404} (May, 2010) {926--930}},
  [\href{http://arxiv.org/abs/arXiv:0905.3176}{{\tt arXiv:0905.3176}}].

\bibitem{Massey2013}
R.~{Massey}, H.~{Hoekstra}, T.~{Kitching}, J.~{Rhodes}, M.~{Cropper},
  J.~{Amiaux} et~al., \emph{{Origins of weak lensing systematics, and
  requirements on future instrumentation (or knowledge of instrumentation)}},
  \href{http://dx.doi.org/{10.1093/mnras/sts371}}{\emph{Monthly Notices of the
  Royal Astronomical Society} {\bf 429} (February, 2013) {661--678}},
  [\href{http://arxiv.org/abs/arXiv:1210.7690}{{\tt arXiv:1210.7690}}].

\bibitem{Hildebrandt2017}
H.~{Hildebrandt}, M.~{Viola}, C.~{Heymans}, S.~{Joudaki}, K.~{Kuijken},
  C.~{Blake} et~al., \emph{{KiDS-450: cosmological parameter constraints from
  tomographic weak gravitational lensing}},
  \href{http://dx.doi.org/{10.1093/mnras/stw2805}}{\emph{Monthly Notices of the
  Royal Astronomical Society} {\bf 465} (February, 2017) 1454--1498},
  [\href{http://arxiv.org/abs/arXiv:1606.05338}{{\tt arXiv:1606.05338}}].

\bibitem{Refregier2003a}
A.~{Refregier}, \emph{{Shapelets - I. A method for image analysis}},
  \href{http://dx.doi.org/{10.1046/j.1365-8711.2003.05901.x}}{\emph{Monthly
  Notices of the Royal Astronomical Society} {\bf 338} (January, 2003)
  {35--47}}, [\href{http://arxiv.org/abs/astro-ph/0105178}{{\tt
  astro-ph/0105178}}].

\bibitem{Zuntz2017}
J.~{Zuntz}, E.~{Sheldon}, S.~{Samuroff}, M.~{Troxel}, M.~{Jarvis},
  N.~{MacCrann} et~al., \emph{{Dark Energy Survey Year 1 Results: Weak Lensing
  Shape Catalogues}},  \href{http://arxiv.org/abs/arXiv:1708.01533}{{\tt
  arXiv:1708.01533}}.

\bibitem{Bertin2011}
E.~{Bertin}, \emph{{Automated Morphometry with SExtractor and PSFEx}},  in
  \emph{Astronomical Data Analysis Software and Systems XX} (I.~{Evans},
  A.~{Accomazzi}, D.~{Mink} and A.~{Rots}, eds.), vol.~435 of
  \emph{Astronomical Society of the Pacific Conference Series}, July, 2011.

\bibitem{Jollife2016}
I.~T. {Jollife} and J.~{Cadima}, \emph{{Principal component analysis: a review
  and recent developments}},
  \href{http://dx.doi.org/{10.1098/rsta.2015.0202}}{\emph{{Philosophical
  Transactions of the Royal Society of London A: Mathematical, Physical and
  Engineering Sciences}} {\bf 374} (2016) }.

\bibitem{Kitching2013}
T.~{Kitching}, B.~{Rowe}, M.~{Gill}, C.~{Heymans}, R.~{Massey}, D.~{Witherick}
  et~al., \emph{{Image Analysis for Cosmology: Results from the GREAT10 Star
  Challenge}}, \href{http://dx.doi.org/{10.1088/0067-0049/205/2/12}}{\emph{The
  Astrophysical Journal Supplement Series} {\bf 205} (April, 2013) 12},
  [\href{http://arxiv.org/abs/arXiv:1210.1979}{{\tt arXiv:1210.1979}}].

\bibitem{Refregier2014}
A.~{Refregier} and A.~{Amara}, \emph{{A way forward for Cosmic Shear:
  Monte-Carlo Control Loops}},
  \href{http://dx.doi.org/10.1016/j.dark.2014.01.002}{\emph{Physics of the Dark
  Universe} {\bf 3} (April, 2014) 1--3},
  [\href{http://arxiv.org/abs/arXiv:1303.4739}{{\tt arXiv:1303.4739}}].

\bibitem{Bruderer2016}
C.~{Bruderer}, C.~{Chang}, A.~{Refregier}, A.~{Amara}, J.~{Berg{\'{e}}} and
  L.~{Gamper}, \emph{{Calibrated Ultra Fast Image Simulations for the Dark
  Energy Survey}},
  \href{http://dx.doi.org/10.3847/0004-637X/817/1/25}{\emph{The Astrophysical
  Journal} {\bf 817} (January, 2016) {25}},
  [\href{http://arxiv.org/abs/arXiv:1504.02778}{{\tt arXiv:1504.02778}}].

\bibitem{Herbel2017}
J.~{Herbel}, T.~{Kacprzak}, A.~{Amara}, A.~{Refregier}, C.~{Bruderer} and
  A.~{Nicola}, \emph{{The redshift distribution of cosmological samples: a
  forward modeling approach}},
  \href{http://dx.doi.org/{10.1088/1475-7516/2017/08/035}}{\emph{Journal of
  Cosmology and Astroparticle Physics} {\bf 8} (August, 2017) {035}},
  [\href{http://arxiv.org/abs/arXiv:1705.05386}{{\tt arXiv:1705.05386}}].

\bibitem{Bruderer2017}
C.~{Bruderer}, A.~{Nicola}, A.~{Amara}, A.~{Refregier}, J.~{Herbel} and
  T.~{Kacprzak}, \emph{{Cosmic shear calibration with forward modeling}},
  \href{http://arxiv.org/abs/arXiv:1707.06233}{{\tt arXiv:1707.06233}}.

\bibitem{LeCun2015}
Y.~{LeCun}, Y.~{Bengio} and G.~{Hinton}, \emph{{Deep Learning}},
  \href{http://dx.doi.org/10.1038/nature14539}{\emph{Nature} {\bf 521} (May,
  2015) 436--444}.

\bibitem{Ravanbakhsh2016}
S.~{Ravanbakhsh}, F.~{Lanusse}, R.~{Mandelbaum}, J.~{Schneider} and
  B.~{P{\'o}czos}, \emph{{Enabling Dark Energy Science with Deep Generative
  Models of Galaxy Images}},  \href{http://arxiv.org/abs/arXiv:1609.05796}{{\tt
  arXiv:1609.05796}}.

\bibitem{Charnock2017}
T.~{Charnock} and A.~{Moss}, \emph{{Deep Recurrent Neural Networks for
  Supernovae Classification}},
  \href{http://dx.doi.org/{10.3847/2041-8213/aa603d}}{\emph{The Astrophysical
  Journal Letters} {\bf 837} (March, 2017) {L28}},
  [\href{http://arxiv.org/abs/arXiv:1606.07442}{{\tt arXiv:1606.07442}}].

\bibitem{Hezaveh2017}
Y.~D. {Hezaveh}, L.~P. {Levasseur} and P.~J. {Marshall}, \emph{{Fast automated
  analysis of strong gravitational lenses with convolutional neural networks}},
  \href{http://dx.doi.org/{10.1038/nature23463}}{\emph{Nature} {\bf 548}
  (August, 2017) {555--557}},
  [\href{http://arxiv.org/abs/arXiv:1708.08842}{{\tt arXiv:1708.08842}}].

\bibitem{Schmelzle2017}
J.~{Schmelzle}, A.~{Lucchi}, T.~{Kacprzak}, A.~{Amara}, R.~{Sgier},
  A.~{R{\'{e}}fr{\'{e}}gier} et~al., \emph{{Cosmological model discrimination
  with Deep Learning}},  \href{http://arxiv.org/abs/arXiv:1707.05167}{{\tt
  arXiv:1707.05167}}.

\bibitem{Lanusse2018}
F.~{Lanusse}, Q.~{Ma}, N.~{Li}, T.~E. {Collett}, C.-L. {Li}, S.~{Ravanbakhsh}
  et~al., \emph{{CMU DeepLens: deep learning for automatic image-based
  galaxy-galaxy strong lens finding}},
  \href{http://dx.doi.org/{10.1093/mnras/stx1665}}{\emph{Monthly Notices of the
  Royal Astronomical Society} {\bf 473} (January, 2018) {3895--3906}},
  [\href{http://arxiv.org/abs/arXiv:1703.02642}{{\tt arXiv:1703.02642}}].

\bibitem{Abolfathi2017}
B.~{Abolfathi}, D.~{Aguado}, G.~{Aguilar}, C.~{Allende Prieto}, A.~{Almeida},
  T.~{Tasnim Ananna} et~al., \emph{{The Fourteenth Data Release of the Sloan
  Digital Sky Survey: First Spectroscopic Data from the extended Baryon
  Oscillation Spectroscopic Survey and from the second phase of the Apache
  Point Observatory Galactic Evolution Experiment}},
  \href{http://arxiv.org/abs/arXiv:1707.09322}{{\tt arXiv:1707.09322}}.

\bibitem{GaiaCollaboration2016a}
{Gaia Collaboration}, \emph{{The Gaia mission}},
  \href{http://dx.doi.org/10.1051/0004-6361/201629272}{\emph{Astronomy {\&}
  Astrophysics} {\bf 595} (November, 2016) {A1}},
  [\href{http://arxiv.org/abs/arXiv:1609.04153}{{\tt arXiv:1609.04153}}].

\bibitem{GaiaCollaboration2016b}
{Gaia Collaboration}, \emph{{Gaia Data Release 1. Summary of the astrometric,
  photometric, and survey properties}},
  \href{http://dx.doi.org/10.1051/0004-6361/201629272}{\emph{Astronomy {\&}
  Astrophysics} {\bf 595} (November, 2016) {A2}},
  [\href{http://arxiv.org/abs/arXiv:1609.04172}{{\tt arXiv:1609.04172}}].

\bibitem{Press2007}
W.~H. {Press}, S.~A. {Teukolsky}, W.~T. {Vetterling} and B.~P. {Flannery},
  \emph{{Numerical Recipes: The Art of Scientific Computing}}.
\newblock Cambridge University Press, 3.~ed., September, 2007.

\bibitem{Massey2005}
R.~{Massey} and A.~{Refregier}, \emph{{Polar shapelets}},
  \href{http://dx.doi.org/{10.1111/j.1365-2966.2005.09453.x}}{\emph{{Monthly
  Notices of the Royal Astronomical Society}} {\bf {363}} ({October}, {2005})
  {197--210}}, [\href{http://arxiv.org/abs/astro-ph/0408445}{{\tt
  astro-ph/0408445}}].

\bibitem{Moffat1969}
A.~{Moffat}, \emph{{A Theoretical Investigation of Focal Stellar Images in the
  Photographic Emulsion and Application to Photographic Photometry}},
  {\emph{Astronomy {\&} Astrophysics} {\bf 3} (December, 1969) 455}.

\bibitem{Bacon2006}
D.~{Bacon}, D.~{Goldberg}, B.~{Rowe} and A.~{Taylor}, \emph{{Weak gravitational
  flexion}},
  \href{http://dx.doi.org/10.1111/j.1365-2966.2005.09624.x}{\emph{{Monthly
  Notices of the Royal Astronomical Society}} {\bf 365} (January, 2006)
  414--428}, [\href{http://arxiv.org/abs/astro-ph/0504478}{{\tt
  astro-ph/0504478}}].

\bibitem{Rhodes2000}
J.~{Rhodes}, A.~{Refregier} and E.~J. {Groth}, \emph{{Weak Lensing
  Measurements: A Revisited Method and Application to Hubble Space Telescope
  Images}}, \href{http://dx.doi.org/{10.1086/308902}}{\emph{{The Astrophysical
  Journal}} {\bf 536} (June, 2000) 79--100},
  [\href{http://arxiv.org/abs/astro-ph/9905090}{{\tt astro-ph/9905090}}].

\bibitem{Berge2013}
J.~{Berg{\'{e}}}, L.~{Gamper}, A.~{R{\'{e}}fr{\'{e}}gier} and A.~{Amara},
  \emph{{An Ultra Fast Image Generator (\textsc{UFig}) for wide-field
  astronomy}},
  \href{http://dx.doi.org/10.1016/j.ascom.2013.01.001}{\emph{Astronomy and
  Computing} {\bf 1} (February, 2013) 23--32},
  [\href{http://arxiv.org/abs/arXiv:1209.1200}{{\tt arXiv:1209.1200}}].

\bibitem{Akeret2013}
J.~{Akeret}, S.~{Seehars}, A.~{Amara}, A.~{Refregier} and A.~{Csillaghy},
  \emph{{CosmoHammer: Cosmological parameter estimation with the MCMC Hammer}},
  \href{http://dx.doi.org/10.1016/j.ascom.2013.06.003}{\emph{Astronomy and
  Computing} {\bf 2} (August, 2013) 27--39},
  [\href{http://arxiv.org/abs/arXiv:1212.1721}{{\tt arXiv:1212.1721}}].

\bibitem{Murphy2012}
K.~P. {Murphy}, \emph{{Machine Learning: a Probabilistic Perspective}}.
\newblock MIT Press, Cambridge, MA, USA, 2012.

\bibitem{Goodfellow2016}
I.~{Goodfellow}, Y.~{Bengio} and A.~{Courville}, \emph{{Deep Learning}}.
\newblock MIT Press, Cambridge, MA, USA, 2016.

\bibitem{Rawat2017}
W.~{Rawat} and Z.~{Wang}, \emph{{Deep Convolutional Neural Networks for Image
  Classification: A Comprehensive Review}},
  \href{http://dx.doi.org/10.1162/neco\_a\_00990}{\emph{{Neural Computation}}
  {\bf 29} (September, 2017) 2352--2449}.

\bibitem{Liu2017}
Q.~{Liu}, N.~{Zhang}, W.~{Yang}, S.~{Wang}, Z.~{Cui}, X.~{Chen} et~al.,
  \emph{{A Review of Image Recognition with Deep Convolutional Neural
  Network}}, pp.~69--80.
\newblock Springer International Publishing, Cham, 2017.
\newblock {10.1007/978-3-319-63309-1\_7}.

\bibitem{McCann2017}
M.~T. {McCann}, K.~H. {Jin} and M.~{Unser}, \emph{{Convolutional Neural
  Networks for Inverse Problems in Imaging: A Review}},
  \href{http://dx.doi.org/10.1109/MSP.2017.2739299}{\emph{IEEE Signal
  Processing Magazine} {\bf 34} (November, 2017) 85--95}.

\bibitem{Krizhevsky2012}
A.~{Krizhevsky}, I.~{Sutskever} and G.~E. {Hinton}, \emph{{ImageNet
  Classification with Deep Convolutional Neural Networks}},  in
  \emph{{Proceedings of the 25th International Conference on Neural Information
  Processing Systems - Volume 1}}, NIPS'12, (USA), pp.~1097--1105, Curran
  Associates Inc., 2012.

\bibitem{Simonyan2014}
K.~{Simonyan} and Z.~Andrew, \emph{{Very Deep Convolutional Networks for
  Large-Scale Image Recognition}},
  \href{http://arxiv.org/abs/arXiv:1409.1556}{{\tt arXiv:1409.1556}}.

\bibitem{Glorot2011}
X.~{Glorot}, A.~{Bordes} and Y.~{Bengio}, \emph{{Deep Sparse Rectifier Neural
  Networks}},  in \emph{Proceedings of the Fourteenth International Conference
  on Artificial Intelligence and Statistics} (G.~{Gordon}, D.~{Dunson} and
  M.~{Dud{\'\i}k}, eds.), vol.~15 of \emph{Proceedings of Machine Learning
  Research}, (Fort Lauderdale, FL, USA), pp.~315--323, PMLR, April, 2011.

\bibitem{Abadi2016}
M.~{Abadi}, A.~{Agarwal}, P.~{Barham}, E.~{Brevdo}, Z.~{Chen}, C.~{Citro}
  et~al., \emph{{TensorFlow: Large-Scale Machine Learning on Heterogeneous
  Distributed Systems}},  \href{http://arxiv.org/abs/arXiv:1603.04467}{{\tt
  arXiv:1603.04467}}.

\bibitem{Huber1964}
P.~J. {Huber}, \emph{{Robust Estimation of a Location Parameter}},
  \href{http://dx.doi.org/{10.1214/aoms/1177703732}}{\emph{The Annals of
  Mathematical Statistics} {\bf 35} (March, 1964) 73--101}.

\bibitem{Ruder2016}
S.~{Ruder}, \emph{{An overview of gradient descent optimization algorithms}},
  \href{http://arxiv.org/abs/arXiv:1609.04747}{{\tt arXiv:1609.04747}}.

\bibitem{Kingma2014}
D.~P. {Kingma} and J.~{Ba}, \emph{{Adam: A Method for Stochastic
  Optimization}},  \href{http://arxiv.org/abs/arXiv:1412.6980}{{\tt
  arXiv:1412.6980}}.

\bibitem{Fischler1981}
M.~A. {Fischler} and R.~C. {Bolles}, \emph{{Random sample consensus: a paradigm
  for model fitting with applications to image analysis and automated
  cartography}},
  \href{http://dx.doi.org/{10.1145/358669.358692}}{\emph{Communications of the
  ACM} {\bf 24} (June, 1981) 381--395}.

\bibitem{Pedregosa2011}
F.~{Pedregosa}, G.~{Varoquaux}, A.~{Gramfort}, V.~{Michel}, B.~{Thirion},
  O.~{Grisel} et~al., \emph{{Scikit-learn: Machine Learning in Python}},
  {\emph{Journal of Machine Learning Research} {\bf 12} (October, 2011)
  {2825--2830}}, [\href{http://arxiv.org/abs/arXiv:1201.0490}{{\tt
  arXiv:1201.0490}}].

\bibitem{Jones2001}
E.~{Jones}, T.~{Oliphant}, P.~{Peterson} et~al., \emph{{SciPy}: Open source
  scientific tools for {Python}},  2001--.

\bibitem{Annis2014}
J.~{Annis}, M.~{Soares-Santos}, M.~A. {Strauss}, A.~C. {Becker}, S.~{Dodelson},
  X.~{Fan} et~al., \emph{{The Sloan Digital Sky Survey Coadd: 275 deg$^{2}$ of
  Deep Sloan Digital Sky Survey Imaging on Stripe 82}},
  \href{http://dx.doi.org/{10.1088/0004-637X/794/2/120}}{\emph{The
  Astrophysical Journal} {\bf 794} (October, 2014) {120}},
  [\href{http://arxiv.org/abs/arXiv:1111.6619}{{\tt arXiv:1111.6619}}].

\bibitem{Nicola2016}
A.~{Nicola}, A.~{Refregier} and A.~{Amara}, \emph{{Integrated approach to
  cosmology: Combining CMB, large-scale structure, and weak lensing}},
  \href{http://dx.doi.org/{10.1103/PhysRevD.94.083517}}{\emph{Physical Review
  D} {\bf 94} (October, 2016) {083517}},
  [\href{http://arxiv.org/abs/arXiv:1607.01014}{{\tt arXiv:1607.01014}}].

\bibitem{Amara2007}
A.~{Amara} and A.~{R{\'{e}}fr{\'{e}}gier}, \emph{{Optimal surveys for
  weak-lensing tomography}}, {\emph{Monthly Notices of the Royal Astronomical
  Society} {\bf 381} (2007) {1018--1026}},
  [\href{http://arxiv.org/abs/astro-ph/0610127}{{\tt astro-ph/0610127}}].

\bibitem{vanderWalt2011}
S.~{van der Walt}, S.~C. {Colbert } and G.~{Varoquaux}, \emph{{The NumPy Array:
  A Structure for Efficient Numerical Computation}},
  \href{http://dx.doi.org/{10.1109/MCSE.2011.37}}{\emph{Computing In Science \&
  Engineering} {\bf 13} (March, 2011) {22--30}},
  [\href{http://arxiv.org/abs/arXiv:1102.1523}{{\tt arXiv:1102.1523}}].

\bibitem{Hunter2007}
J.~D. {Hunter}, \emph{{Matplotlib: A 2D graphics environment}},
  \href{http://dx.doi.org/{10.1109/MCSE.2007.55}}{\emph{Computing In Science \&
  Engineering} {\bf 9} (June, 2007) {90--95}}.

\bibitem{Waskom2017}
M.~{Waskom}, O.~{Botvinnik}, D.~{O'Kane}, P.~{Hobson}, S.~{Lukauskas}, D.~C.
  {Gemperline} et~al., \emph{{mwaskom/seaborn: v0.8.1 (September 2017)}},
  September, 2017.
\newblock {10.5281/zenodo.883859}.

\bibitem{Kluyver2016}
T.~{Kluyver}, B.~{Ragan-Kelley}, F.~{P{\'e}rez}, B.~{Granger}, M.~{Bussonnier},
  J.~{Frederic} et~al., \emph{{Jupyter Notebooks -- a publishing format for
  reproducible computational workflows}},  in \emph{{Positioning and Power in
  Academic Publishing: Players, Agents and Agendas}} ({F. Loizides and B.
  Schmidt}, ed.), pp.~{87--90}, {IOS Press}, 2016.

\bibitem{Perez2007}
F.~{P{\'e}rez} and B.~E. {Granger}, \emph{{{IP}ython: a System for Interactive
  Scientific Computing}},
  \href{http://dx.doi.org/{10.1109/MCSE.2007.53}}{\emph{Computing in Science
  and Engineering} {\bf 9} (May, 2007) {21--29}}.

\bibitem{AstropyCollaboration2018}
{The Astropy Collaboration}, A.~{Price-Whelan}, B.~{Sip{\H o}cz},
  H.~{G{\"u}nther}, P.~{Lim}, S.~{Crawford} et~al., \emph{{The Astropy Project:
  Building an inclusive, open-science project and status of the v2.0 core
  package}},  \href{http://arxiv.org/abs/arXiv:1801.02634}{{\tt
  arXiv:1801.02634}}.

\bibitem{Lam2015}
S.~K. {Lam}, A.~{Pitrou} and S.~{Seibert}, \emph{{Numba: A LLVM-based Python
  JIT Compiler}},  in \emph{{Proceedings of the Second Workshop on the LLVM
  Compiler Infrastructure in HPC}}, LLVM '15, (New York, NY, USA),
  pp.~{7:1--7:6}, ACM, 2015.
\newblock \href{http://dx.doi.org/{10.1145/2833157.2833162}}{DOI}.

\bibitem{McKinney2011}
W.~{McKinney}, \emph{pandas: a foundational python library for data analysis
  and statistics}, .

\end{thebibliography}\endgroup

\appendix

\section{SDSS CasJobs SQL query}
\label{app:casjobs-sql-query}

Here, we give the SQL queries used to obtain the list of SDSS images from which we extracted our star sample described in section \ref{sec:sdss-star-sample}. We ran these queries on the SDSS data release 14.

\begin{enumerate}

\item \texttt{SELECT DISTINCT fieldID FROM PhotoObjAll
INTO mydb.field\_id
WHERE \\
PhotoObjAll.clean = 1 \\
AND PhotoObjAll.type\_r = 6 \\
AND PhotoObjAll.psfFlux\_r * SQRT(PhotoObjAll.psfFluxIvar\_r) > 50}

\item \texttt{SELECT ALL Field.run, Field.field, Field.camcol, Field.ra, Field.dec, \\
Field.gain\_r, Field.nMgyPerCount\_r, Field.sky\_r, Field.skySig\_r, \\
Field.pixScale \\
INTO mydb.fieldlist \\
FROM Field INNER JOIN mydb.field\_id ON Field.fieldID = field\_id.fieldID \\
WHERE Field.ra > 100 \\
AND Field.ra < 280 \\
AND Field.dec > 0 \\
AND Field.dec < 60 \\}
\end{enumerate}

\section{SVD of the modeled SDSS sample}
\label{app:svd-cnn}

\begin{figure}[htb]
\centering
\includegraphics[width=\textwidth]{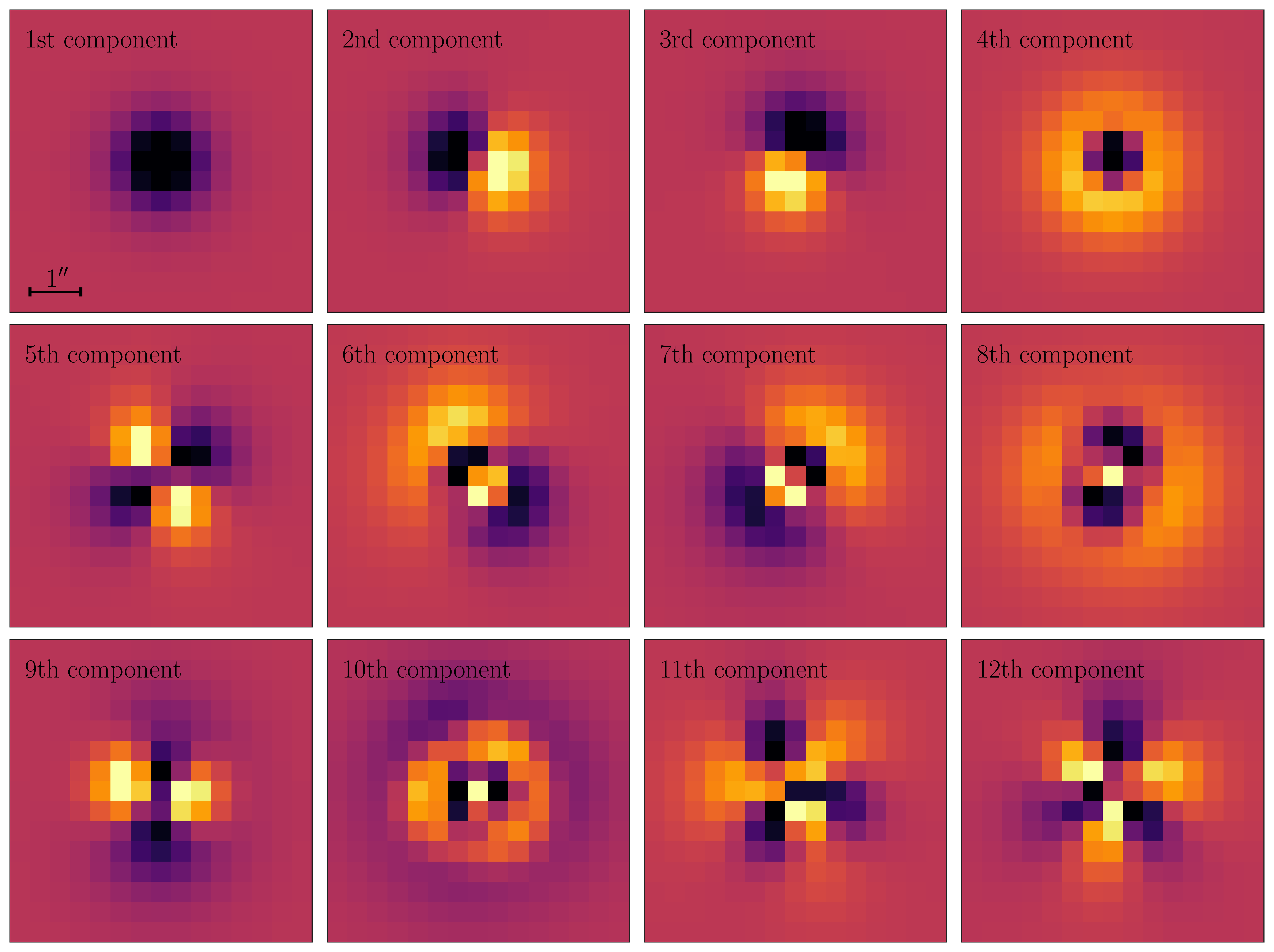}
\caption{\label{fig:svd-cnn}Uncentered principal components of the model images of the SDSS stars described in section \ref{sec:results}. The similarity of this decomposition and the one of the SDSS sample, which is given in figure \ref{fig:svd-sdss}, further validates our method. The color scale is the same for both figures. As explained in the text above, sign flips are not important here, since they cancel out when decomposing a PSF image into its principal components.} 
\end{figure}

In this appendix, we display the uncentered principal components of the modeled SDSS stars (figure \ref{fig:svd-cnn}), which are also referenced in the main text. Note that sign flips (visible e.g. in the first component when comparing this figure to figure \ref{fig:svd-sdss}) are irrelevant here, because they do not change the space of light distributions spanned by the principal components. When decomposing a PSF image $I$ into principal components $\phi_i$, i.e. $I = \sum_i c_i \phi_i$, the decomposition coefficients $c_i$ are given by  the projections of the image onto the principal components: $c_i = \langle I, \phi_i \rangle$, where $\langle \cdot \rangle$ denotes a scalar product. Thus, any sign flips in the principal components cancel out: $c_i \phi_i = \langle I, \phi_i \rangle \phi_i = \langle I, (-\phi_i) \rangle (-\phi_i)$.

\section{Results from direct fitting}
\label{app:direct-fitting}

\begin{figure}[htb]
\centering
\includegraphics[width=\textwidth]{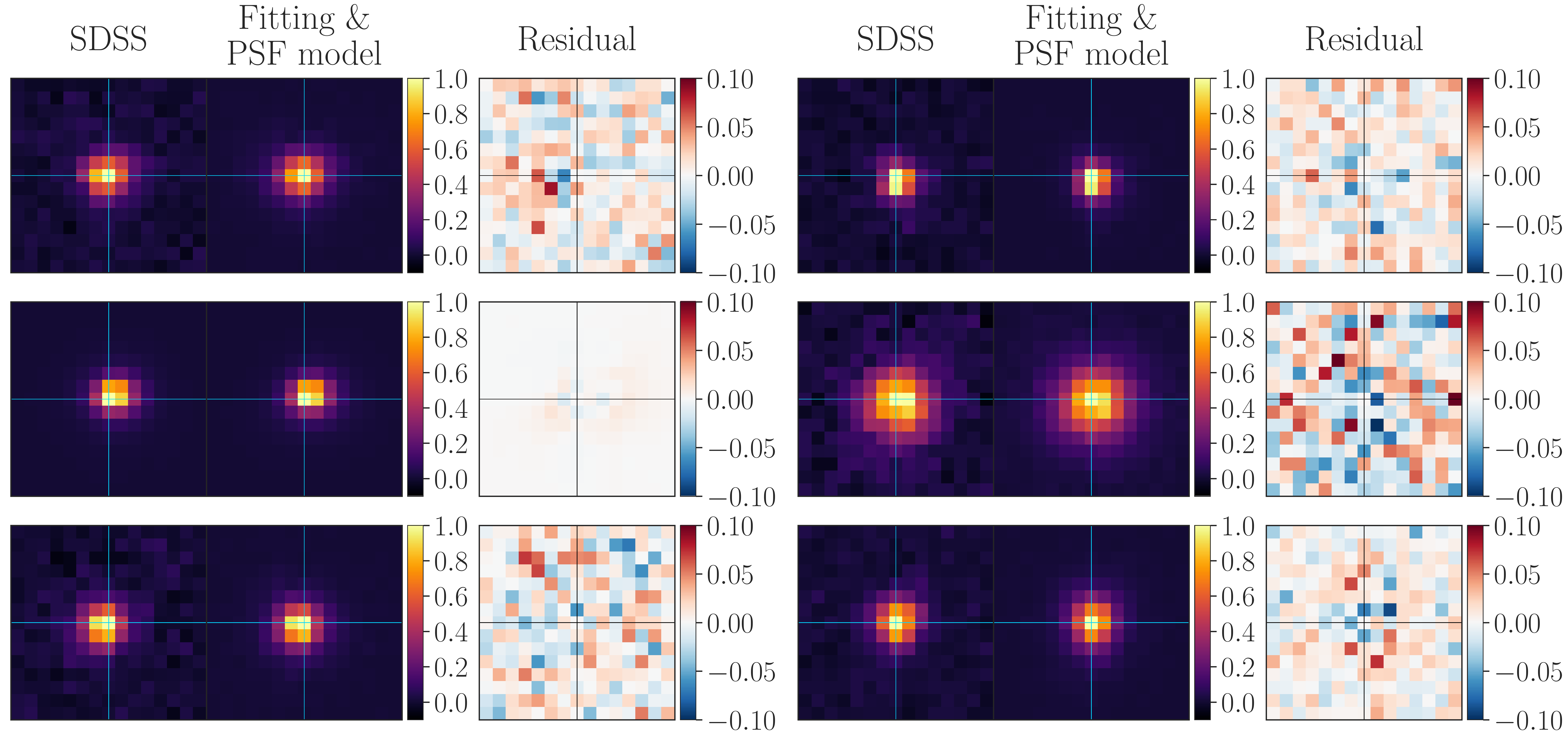}
\caption{\label{fig:sdss-stars--vs--fitting-stars}The figure shows the same six randomly selected stars from the SDSS sample that are also displayed in figure \ref{fig:sdss-stars--vs--cnn-stars}. The difference is that the model images where not generated using  parameters estimates produced by the CNN, instead, the model parameters were obtained by a direct fitting approach. When comparing these results to the ones obtained using the CNN, see figure \ref{fig:sdss-stars--vs--cnn-stars}, it is clear that the direct fitting approach performs similarly well as the CNN in terms of residual pixel values. However, direct fitting has a runtime which is orders of magnitudes larger than the runtime of the network, rendering this method practically unusable for large datasets.}
\end{figure}

Here, we show the results obtained by directly fitting our PSF model to SDSS stars instead of using a CNN, as discussed in section \ref{subsec:comparison-direct-fitting}. We found that the Powell optimization algorithm, implemented within the SciPy library (\texttt{scipy.optimize.minimize}), yields good results in terms of residual pixel values. Figure \ref{fig:sdss-stars--vs--fitting-stars} displays the results obtained by directly fitting the six randomly selected SDSS stars that were also used in figure \ref{fig:sdss-stars--vs--cnn-stars}. As can be seen, direct fitting yields results comparable to the ones obtained using the CNN. However, as discussed in the main text, runtime issues render this method virtually unusable for data volumes produced by currently ongoing and upcoming state-of-the-art wide-field surveys.

\end{document}